\documentclass[12pt,a4paper]{article}
\usepackage[english]{babel}
\usepackage{graphicx,rotating,epsfig,amsmath,multirow,multicol}
\newcommand{\Lumi}{{\cal L}}
\newcommand{\Mchi}{m_{\tilde{\chi}}}
\def\TeV{\ifmmode {\,\mathrm{ Te\kern -0.1em V}}\else
                   \textrm{Te\kern -0.1em V}\fi}%
\def\GeV{\ifmmode {\,\mathrm{ Ge\kern -0.1em V}}\else
                   \textrm{Ge\kern -0.1em V}\fi}%
\let\gev=\GeV

\begin{document}
\begin{flushright}
DESY 05-049\\
18th March 2005
\end{flushright}
\begin{center}
\boldmath
\textbf{\Large Studies on Chargino production and decay at a photon collider}
\unboldmath
\end{center}
\begin{center}
{\textbf{G.Kl\"amke$^{*,}$\footnote{email: klaemke@particle.uni-karlsruhe.de},
K.M{\"o}nig\footnote{email: Klaus.Moenig@desy.de} 
}}
\end{center}
\begin{center}
{Deutsches Elektronen-Synchrotron DESY \\
Platanenallee 6, 15738 Zeuthen, Germany\\
$^*$ Now at Institut f\"ur Theoretische Physik, Universit\"at Karlsruhe
}
\end{center}
%%%%%%%%%%%%%%%%
\begin{abstract}
%%%%%%%%%%%%%%%%
  A Monte-Carlo analysis on production and decay of supersymmetric charginos
  at a future photon-collider is presented. A photon collider offers the
  possibility of a direct branching-ratio measurement. In this study, the
  process $\gamma\gamma \rightarrow
  \tilde{\chi}_1^+\tilde{\chi}_1^-\rightarrow
  W^+W^-\tilde{\chi}_1^0\tilde{\chi}_1^0 \rightarrow
  q\bar{q}q\bar{q}\tilde{\chi}_1^0\tilde{\chi}_1^0$ has been considered for a
  specific mSUGRA scenario. Various backgrounds and a parameterised detector
  simulation have been included. Depending on the centre-of-mass energy, a
  statistical error for the directly measurable branching ratio
  BR(${\tilde{\chi}_1^\pm}\rightarrow {\tilde{\chi}_1^0} W^\pm$) of up to
  3.5\% can be reached.
\end{abstract}

%%%%%%%%%%%%%%%%%%%%%%%%%%%%%%%%%%%%%%%%%%%%%%%%%%%%%%%%%%%%%%%%%%%%%%%
\section{Introduction}
%%%%%%%%%%%%%%%%%%%%%%%%%%%%%%%%%%%%%%%%%%%%%%%%%%%%%%%%%%%%%%%%%%%%%%%

An option for the future Linear Collider project is the
photon collider \cite{Ginzburg:1981ik,Badelek:2001xb}.
Such a collider provides the possibility of studying photon-photon collisions 
up to 80\% of the $e^- e^-$ centre of mass energy. 
If Supersymmetry is realized in nature, then also supersymmetric
particles can be produced and investigated at such a facility. 
The photon collider has the
advantage that the production of charged particle pairs is determined by pure
QED. This offers the possibility to directly measure the decay properties of
supersymmetric particles, once their masses have been precisely measured at
the $e^+e^-$-collider. In addition the production cross sections for charged
particles are significantly larger at a photon collider than in $e^+ e^-$
annihilation. 

In this paper a Monte-Carlo analysis on production and decay of supersymmetric
charginos $\chi^\pm_1$ is presented. The channel $\gamma\gamma \rightarrow
\chi^+_1\chi^-_1 \rightarrow W^+W^-\chi^0_1\chi^0_1\rightarrow
q\bar{q}q\bar{q}\chi^0_1\chi^0_1$ has been studied, where each chargino decays
into a $W^\pm$-boson and a neutralino $\chi^0_1$. The target was to estimate
the statistical error in a direct measurement of the chargino branching ratio
BR(${\tilde{\chi}_1^\pm}\rightarrow {\tilde{\chi}_1^0} W^\pm$). This was done
for a mSUGRA scenario similar to SPS1a \cite{sps} and for two different beam
energies ${\sqrt{s_{ee}}=500 \gev}$ and ${\sqrt{s_{ee}}=600 \gev}$. The main
Standard Model backgrounds and a parameterised detector simulation have been
included. The obtained efficiencies and purities are presented. Finally the
relevance of the photon collider measurements in addition to $e^+e^-$ has been
tested for the precision with which the Supersymmetry breaking parameters in
the MSSM can be obtained.

%%%%%%%%%%%%%%%%%%%%%%%%%%%%%%%%%%%%%%%%%%%%%%%%%%%%%%%%%%%%%%%%%%%%%%%
\section{Choice of a mSUGRA scenario}
%%%%%%%%%%%%%%%%%%%%%%%%%%%%%%%%%%%%%%%%%%%%%%%%%%%%%%%%%%%%%%%%%%%%%%%

A general starting point for the choice of mSUGRA parameters is the SPS1a
scenario \cite{sps}. However, in SPS1a the chargino decays almost entirely
into a stau and a neutrino ${\tilde{\chi}_1^\pm}\rightarrow
{\tilde{\tau}}^\pm_1\nu_\tau$, leaving only a small branching ratio of the
decay $\chi^\pm_1\rightarrow W^\pm \chi^0_1$ \cite{Ghodbane:2002kg}. For this
reason the mSUGRA parameters have been slightly changed for this study in
order to obtain a larger branching ratio for the decay into a $W^\pm$-boson
and a neutralino.  Table~\ref{tbl:supars} shows the chosen values for the
parameters. Only $m_0$ and $\tan \beta$ were modified with respect to SPS1a.
This was done in such a way that $m_{\tilde{\chi}_1^\pm}$ and
$m_{\tilde{\chi}_1^0}$ remained unchanged (Table~\ref{tbl:massbr}). Thus the
kinematical properties of the reaction $\gamma\gamma \rightarrow
\chi^+_1\chi^-_1 \rightarrow W^+W^-\chi^0_1\chi^0_1$ are the same as for the
SPS1a case. However, $m_{{\tilde{\tau}}_1}$ changed as well as the branching
ratio BR(${\tilde{\chi}_1^\pm}\rightarrow {\tilde{\chi}_1^0} W^\pm$) which is
increased from 7\% to 26\%. This has been considered as a more reasonable
number for an analysis of the ${\tilde{\chi}_1^\pm}\rightarrow
{\tilde{\chi}_1^0} W^\pm$ decay.

\begin{table}[htb]
\centering
\renewcommand{\arraystretch}{1.2}
\begin{tabular}{|l||c|c|c|c|c|}
\hline
 Scenario& $m_0$ & $m_{1/2}$ & $A_0$ & $\tan \beta$& sign $\mu$ \\
\hline\hline
SPS1a & $100\gev$ & $250\gev$ & $-100\gev$ & $10$ & $+1$\\
this study& $130\gev$ & $250\gev$ & $-100\gev$ & $9$ & $+1$\\
\hline
\end{tabular}
\caption{The values of the mSUGRA parameters for SPS1a and the scenario used 
in this study.}
\label{tbl:supars}
\end{table}
\begin{table}[hbt]
\centering
\renewcommand{\arraystretch}{1.2}
\begin{tabular}{|c||c|c|}
\hline
Observable & SPS1a & this study\\
\hline\hline
 $m_{\tilde{\chi}_1^\pm}$ & $180.4\gev$ & $180.4\gev$ \\
 $m_{\tilde{\chi}_1^0}$ &$95.6\gev$& $95.6\gev$\\
 $m_{{\tilde{\tau}}_1}$ &$134.4\gev$& $158.8\gev$\\
BR(${\tilde{\chi}_1^\pm}\rightarrow {\tilde{\tau}}^\pm_1 \nu_\tau$) & 91.9\%& 72.4\% \\
BR(${\tilde{\chi}_1^\pm}\rightarrow {\tilde{\chi}_1^0} W^\pm$) & 7.2\% & 26.2\% \\
\hline
\end{tabular}
\caption{Chargino, neutralino and stau masses and the chargino branching 
ratios for SPS1a and the parameter choice used in this study. 
The numbers were calculated with ISAJET~7.67 \cite{Paige:2003mg}.}
\label{tbl:massbr}
\end{table}

%%%%%%%%%%%%%%%%%%%%%%%%%%%%%%%%%%%%%%%%%%%%%%%%%%%%%%%%%%%%%%%%%%%%%%%
\section{The photon collider}
%%%%%%%%%%%%%%%%%%%%%%%%%%%%%%%%%%%%%%%%%%%%%%%%%%%%%%%%%%%%%%%%%%%%%%%

The photon collider ($\gamma\gamma$-collider) is an option for the next Linear
Collider project \cite{Badelek:2001xb}. The idea is to create high energetic
photons by scattering accelerated electrons on a focused laser beam.  For this
purpose the positron beam is replaced by a second $e^-$-beam.  The produced
photon beams allow the study of photon collisions at energies and luminosities
that are comparable to the $e^+e^-$-collider.

\begin{figure}[htb]
\centering
\begin{tabular}{cc}
\includegraphics[width=0.48\linewidth]{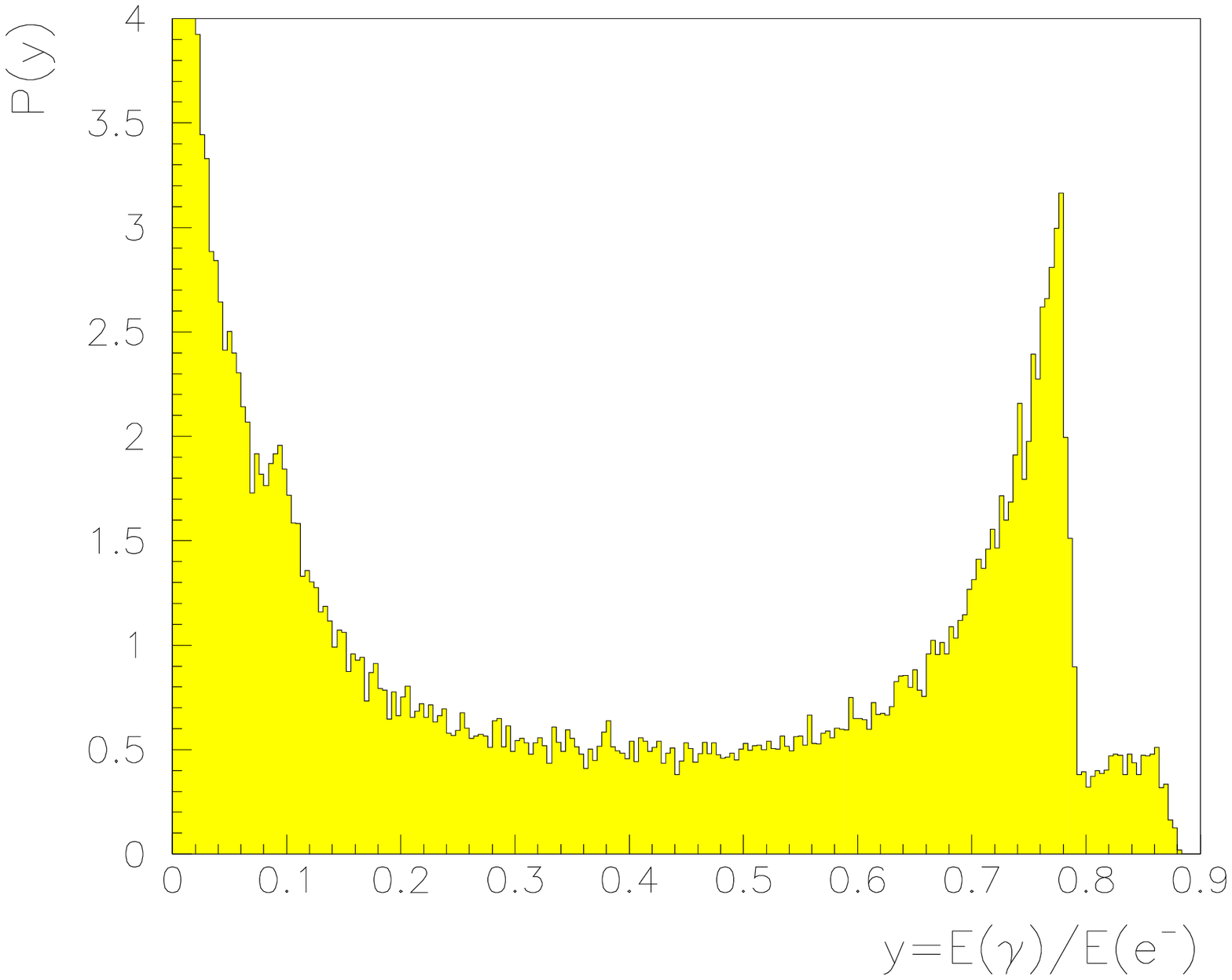} & 
\includegraphics[width=0.48\linewidth]{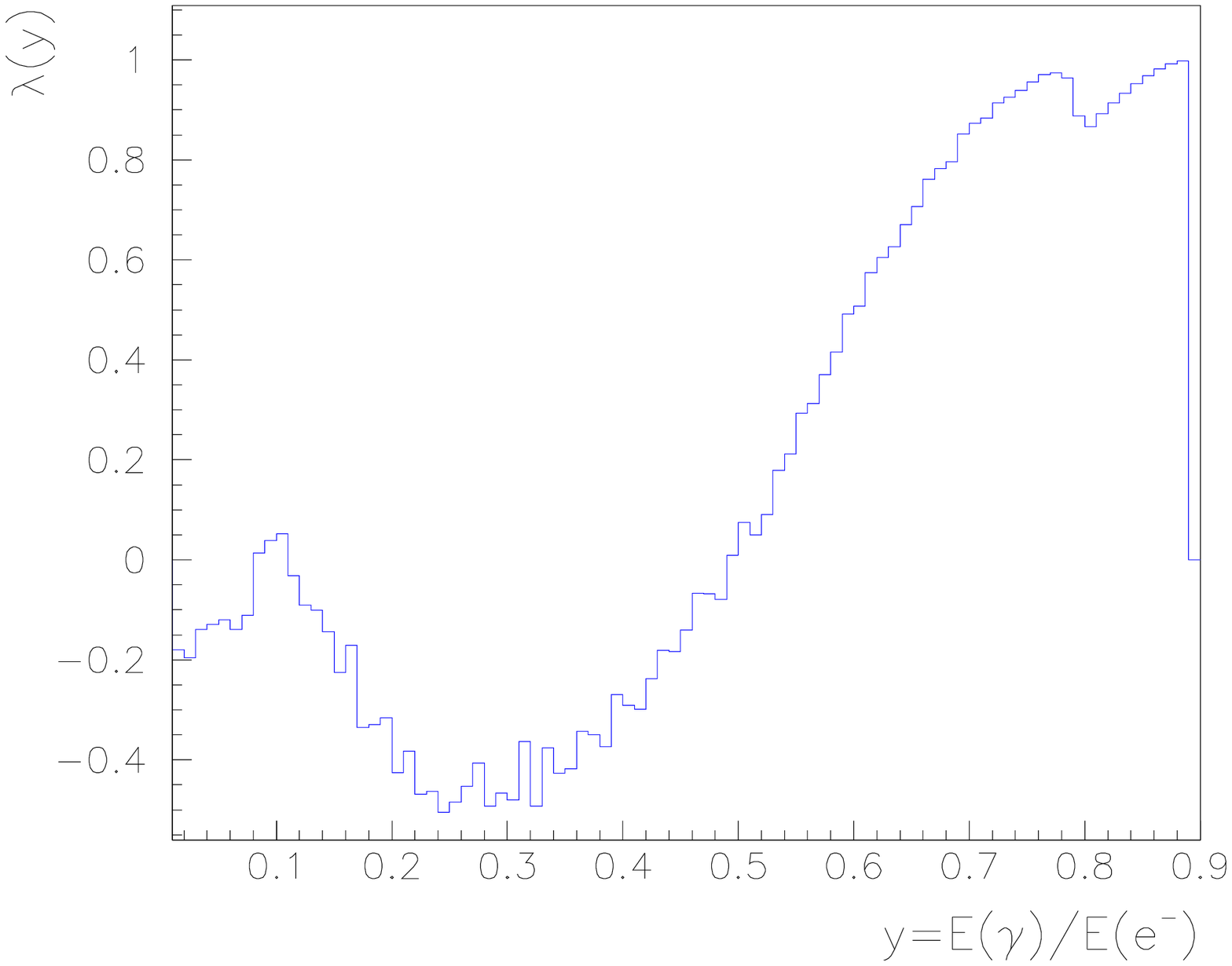}\\
\end{tabular}
\caption{Energy distribution $P(y)$ of the produced photons (left) and 
  photon polarisation $\lambda(y)$ (right) in dependence on $y$, which is the
  ratio of photon energy $E(\gamma)$ and beam-electron energy $E(e^-)$.}
\label{fig:bspec}
\end{figure}

\begin{figure}[hbt]
\centering
\includegraphics[width=0.45\linewidth,bb=4 4 491 473]{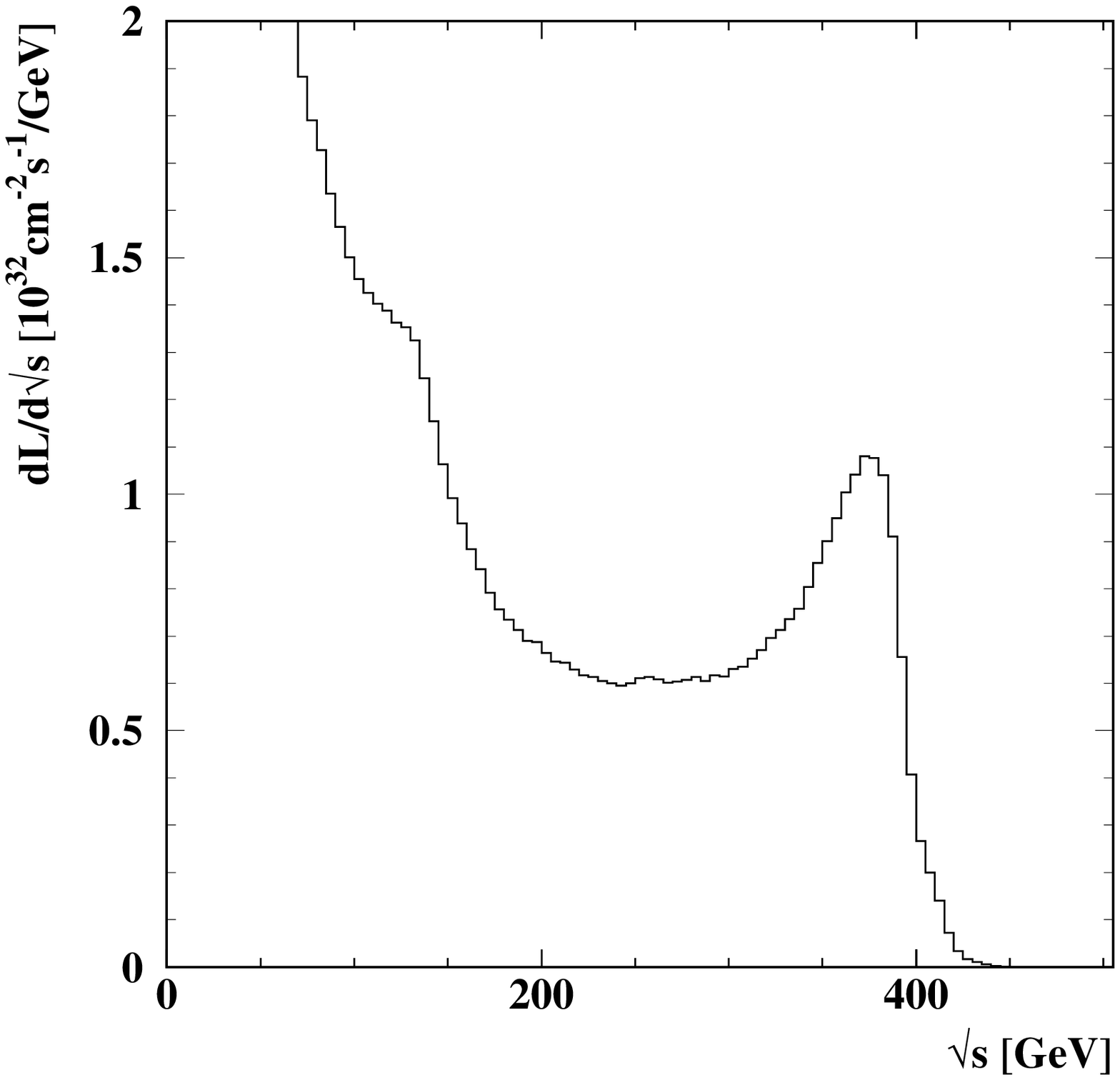}
\includegraphics[width=0.45\linewidth,bb=4 4 491 473]{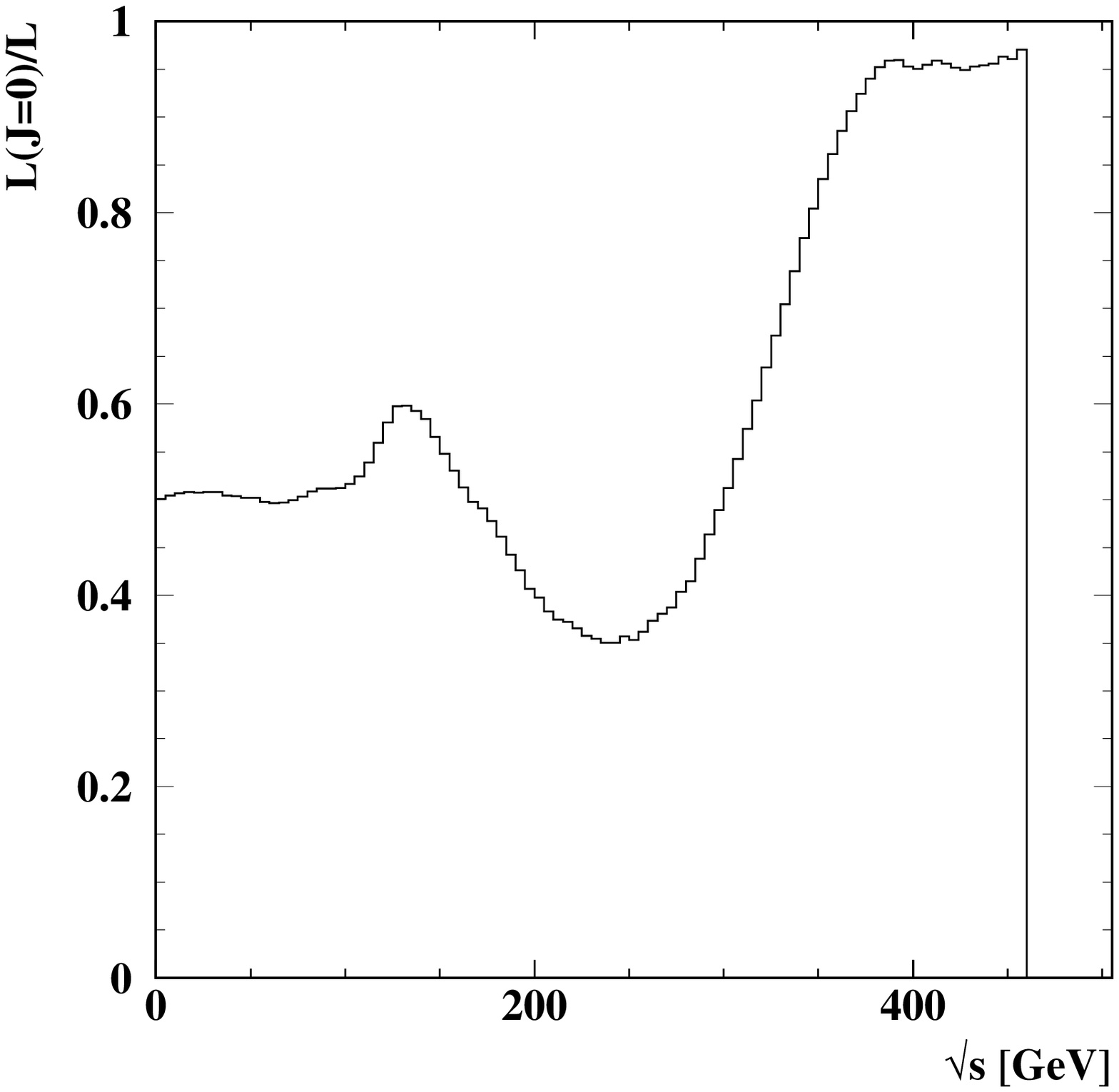}
\caption{ $\gamma\gamma$ luminosity spectrum $d\Lumi/d\sqrt{s_{\gamma\gamma}}$
  (left) and the fraction of the luminosity with $J=0$ (right) in dependence
  of $\sqrt{s_{\gamma\gamma}}$ for a centre-of-mass energy of the two electron
  beams of $\sqrt{s_{ee}}=500 \gev$.}
\label{fig:cain}
\end{figure}

The energy spectrum of the scattered photons is shown in Fig.~\ref{fig:bspec}
(left) for an electron beam energy of $E(e^-)=250 \gev$ \cite{Telnov:2000zx}.
The spectrum is peaked at photon energies of about $70\% - 80\%$ of the
electron energy. The rise at low energies is due to multiple
electron-photon interactions. 
The part of the spectrum above $y\approx0.8E(e^-)$ 
can be explained by nonlinear interactions of an electron with several laser
photons \cite{Badelek:2001xb}. Fig.~\ref{fig:bspec} (right) shows the photon
polarisation spectrum $\lambda(y)$: The high energetic photons are strongly
circular polarised. This can be achieved, by using polarised electron and
laser beams. Here, an electron polarisation of $85\%$ and a laser beam
polarisation of $100\%$ was assumed. 
\begin{figure}[hbt]
\centering
\includegraphics[width=0.8\linewidth]{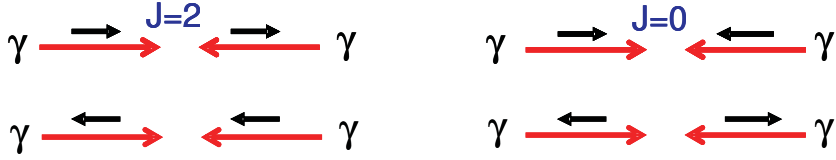}
\caption{The possible alignments of the helicities (short arrows) of the 
  colliding photons that lead to a total angular momentum of $J=2$ or $J=0$.}
\label{fig:helic}
\end{figure} 
The circular polarisation of the photon beams offers two possible running
modes for the $\gamma\gamma$-collider in terms of helicities
(Fig.~\ref{fig:helic}). One with a parallel and one with an anti-parallel
alignment of the photon helicities. These correspond to an overall angular
momentum of either $J=2$ or $J=0$ for the two-photon system.
The luminosity spectrum and the polarisation in dependence of the two-photon
centre-of-mass energy $\sqrt{s_{\gamma\gamma}}$ is shown in
Fig.~\ref{fig:cain}. It has been calculated with the program {\it CAIN}
\cite{Chen:1997fz}. The total luminosity is
$\Lumi_{\gamma\gamma}=10\cdot10^{34}cm^{-2}s^{-1}$ which corresponds to an
integrated luminosity of $1000 fb^{-1}$ per year\footnote{A year is assumed to
  be $10^7$s at design luminosity.}. However the luminosity within the high
energy peak (i.e. $\sqrt{s_{\gamma\gamma}}>300\gev$) is only
$\Lumi_{peak}=1.1\cdot10^{34}cm^{-2}s^{-1}=100 fb^{-1}/\mbox{year}$.

Compared to the $e^+e^-$-collider, a photon collider cannot provide
monochromatic beams. This makes event analyses harder, since the collision
energy, which is important for kinematic constraints, is an unknown variable
here.

%%%%%%%%%%%%%%%%%%%%%%%%%%%%%%%%%%%%%%%%%%%%%%%%%%%%%%%%%%%%%%%%%%%%%%%
\section{Chargino production}
%%%%%%%%%%%%%%%%%%%%%%%%%%%%%%%%%%%%%%%%%%%%%%%%%%%%%%%%%%%%%%%%%%%%%%%

\begin{figure}[htb]
\centering
\includegraphics[width=0.25\linewidth]{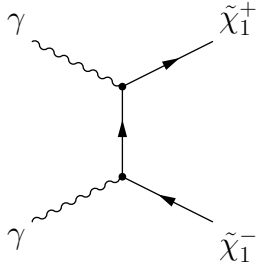}
\caption{Feynman diagram for 
$\gamma\gamma \rightarrow {\tilde{\chi}_1^+}{\tilde{\chi}_1^-}$}
\label{fig:chichiprod}
\end{figure}

The pair production of charginos in photon collisions is described by pure
QED. Fig.~\ref{fig:chichiprod} shows the only leading order diagram for the
$\gamma\gamma \rightarrow {\tilde{\chi}_1^+}{\tilde{\chi}_1^-}$ process.
>From this diagram the total cross section in the centre-of-mass system can be
derived \cite{Mayer:2003sw}:
\begin{equation}
\begin{split}
\sigma_{p,\alpha\beta}=& 
\frac{e^4}{16\pi{E^6}}\biggl\{\left[ \Mchi^2(2E^2-\Mchi^2)+
  2E^4(1-\alpha\beta) \right] \ln{\frac{E+q}{\Mchi}}\\
&+Eq \left[ 2E^2-\Mchi^2-3E^2(1-\alpha\beta) \right] \biggr\}
\end{split}
\label{eqn:chichiprod}
\end{equation}
Where $E$ is the photon beam energy in the centre-of-mass system and
$\alpha$,$\beta$ describe the helicity of the incoming photons. Furthermore
$\Mchi$ and $q=(E^2-\Mchi^2)^{1/2}$ are the chargino mass and momentum and $e$
is the elementary charge.
\begin{figure}[hbt]
\begin{tabular}{cc}
\includegraphics[width=0.47\linewidth,height=0.35\linewidth]{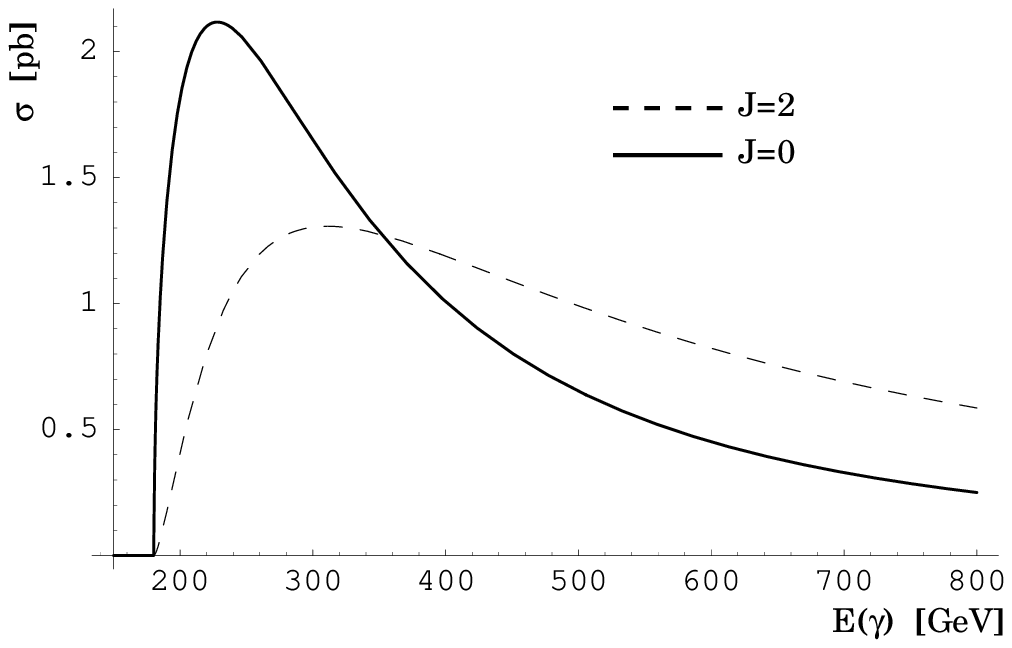}&
\includegraphics[width=0.5\linewidth]{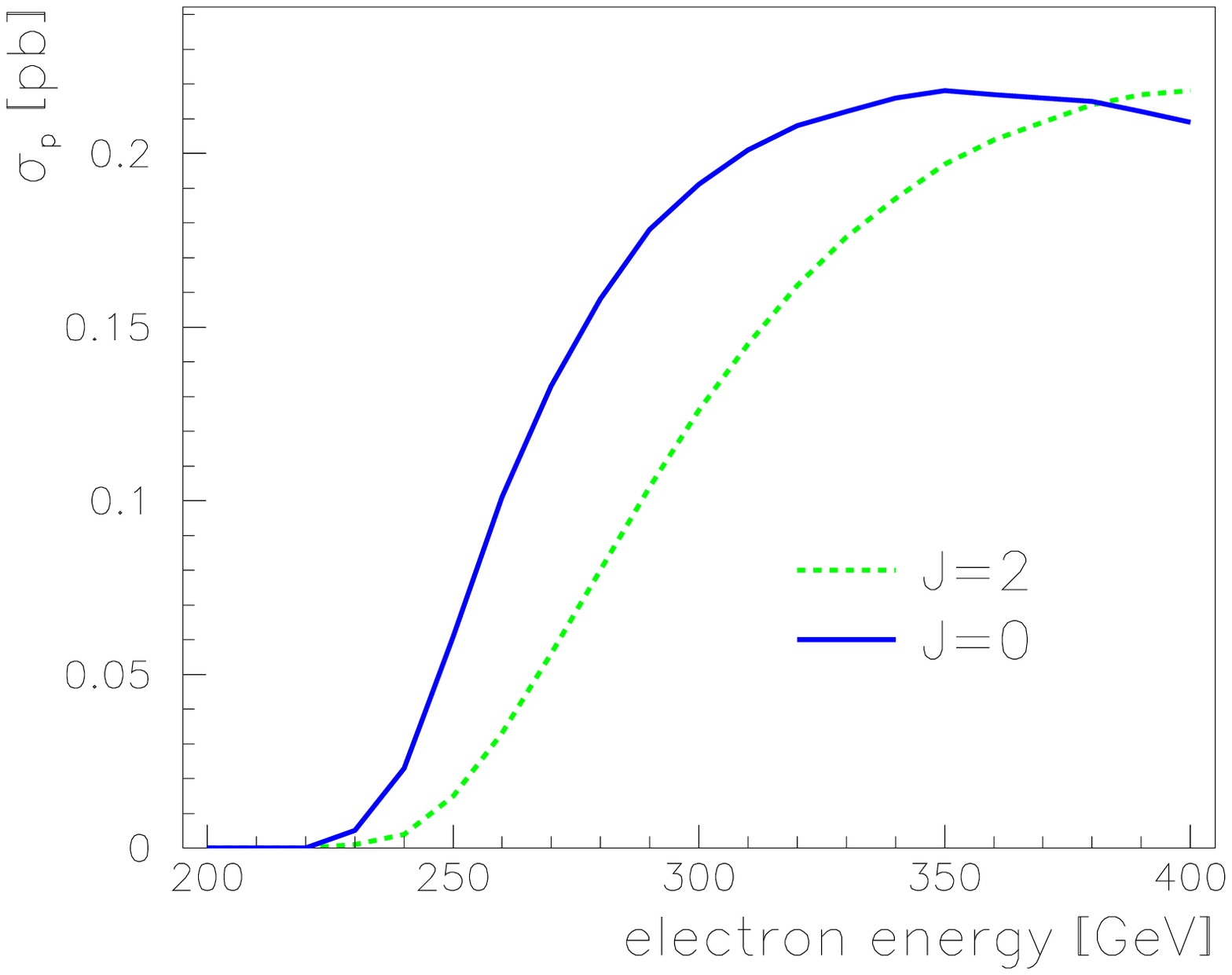}\\
\end{tabular}
\caption{Left: $\sigma_{p,\alpha\beta}$ in dependence of  $E$ for $J=0$ 
  ($\alpha=\beta=\pm 1$) and $J=2$ ($\alpha=-\beta=\pm 1$), $\Mchi=180\gev$.
  Right: effective cross section $\sigma_p$ as a function of the electron beam
  energy $E_{e^-}=\frac{1}{2}\sqrt{s_{ee}}$ for $J=0$ and $J=2$.}
\label{fig:chixs}
\end{figure}
Beside the photon energy and polarisation, the production cross section only
depends on the charge and mass of the chargino. In Fig.~\ref{fig:chixs} (left)
the production cross section is plotted in dependence of the photon energy $E$
for the $J=2$ and $J=0$ mode. Because of parity conservation only the product
$\alpha \cdot \beta =\pm 1$ is relevant. For energies less than $350 \gev$
especially near the production threshold ($E=\Mchi=180 \gev$) the cross
section is larger for the $J=0$ mode, while this behaviour flips for higher
energies. The maximum cross section is $\sigma\approx 2.1$ pb at $E\approx 230
\gev$.  

At a photon collider there are no monochromatic photon beams with
fixed energy. The photons spread over a wide energy range. Thus the production
cross section has to be convoluted with the luminosity spectrum
$d\Lumi/d\sqrt{s_{\gamma\gamma}}$ and the polarisation spectrum $\lambda(y)$
\cite{Mayer:2003sw}:
\begin{equation}
\sigma_p(s_{\gamma\gamma})=\frac{1}{4}\sum_{\alpha,\beta=\pm1}[1+\alpha\lambda(y_1)][1+\beta\lambda(y_2)]\sigma_{p,\alpha\beta}(s_{\gamma\gamma})
\label{eqn:convpol}
\end{equation}
\begin{equation}
\sigma_p(s_{ee})=\int{d\Lumi/d\sqrt{s_{\gamma\gamma}}\sigma_p(s_{\gamma\gamma}=y_1y_2s_{ee})d\sqrt{s_{\gamma\gamma}}}
\label{eqn:convener}
\end{equation} 
Equation~\ref{eqn:convpol} describes the weighting of the cross section
$\sigma_{p,\alpha\beta}$ with the mean helicities $\lambda(y_1)$,
$\lambda(y_2)$ of the incoming photons. The resulting cross section
$\sigma_p(s_{\gamma\gamma})$ is convoluted with the luminosity spectrum 
(eqn.~\ref{eqn:convener}). One obtains an effective production cross section
$\sigma_p(s_{ee})$ for the overall process \mbox{$e^-e^- \rightarrow
  \gamma\gamma \rightarrow {\tilde{\chi}_1^+}{\tilde{\chi}_1^-}$} in the
$e^-e^-$ centre-of-mass system which is plotted in Fig.~\ref{fig:chixs}
(right) for the two different helicity modes $J=0,2$. It has been calculated
with {\it SHERPA} \cite{Gleisberg:2003xi}. For beam energies below $380 \gev$
the $J=0$ configuration provides the larger cross section, therefore that mode
is used in the following for this analysis. 
In the region, where the $J=0$ and $J=2$ mode are similar, we expect
similar results for both modes. However the $J=2$ mode has not been
studied in detail.
In general the effective cross
section is clearly smaller than the cross section for monochromatic beams.
This is due to the fact that a major part of the colliding photons have too
little energy to fulfil the threshold condition
$s_{\gamma\gamma}=y_1y_2s_{ee}>(2\Mchi)^2$.
\begin{table}[hbt]
\centering
\renewcommand{\arraystretch}{1.4}
\begin{tabular}{|c|c|c|}
\hline
$\sqrt{s_{ee}}=500$ GeV  & $\sigma_p=64.7$ fb & $\approx64.7\cdot10^3$  ${\tilde{\chi}_1^+}{\tilde{\chi}_1^-}$ - pairs / year ($10^7s$)\\
$\sqrt{s_{ee}}=600$ GeV  & $\sigma_p=198.0$ fb & $\approx198\cdot10^3$  ${\tilde{\chi}_1^+}{\tilde{\chi}_1^-}$ - pairs / year ($10^7s$)\\
\hline
\end{tabular}
\caption{Values for the effective cross section $\sigma_p$ and the number of 
produced ${\tilde{\chi}_1^+}{\tilde{\chi}_1^-}$-pairs per year for $J=0$.}
\label{tbl:chipairs}
\end{table}
It should be stressed that this effective cross section is not a cross
section in the conventional sense, since it implicitly contains information
about the luminosity spectrum. In order to obtain the number of produced
chargino pairs per year, $\sigma_p(s_{ee})$ has to be multiplied with the
integrated photon luminosity of $\Lumi_{\gamma\gamma}^{int}=1000 fb^{-1}$.
This leads to $\approx64.7\cdot10^3$ chargino pairs per year for a beam energy
of $E_{e^-}=250 \gev$ (i.e. $\sqrt{s_{ee}}=500 \GeV$) and
$\approx198\cdot10^3$ pairs for $\sqrt{s_{ee}}=600 \GeV$ 
(Table~\ref{tbl:chipairs}). So at $600 \gev$ there are about three times more
produced chargino pairs than for $500 \gev$.

%%%%%%%%%%%%%%%%%%%%%%%%%%%%%%%%%%%%%%%%%%%%%%%%%%%%%%%%%%%%%%%%%%%%%%%
\section{Signal and background simulation}
%%%%%%%%%%%%%%%%%%%%%%%%%%%%%%%%%%%%%%%%%%%%%%%%%%%%%%%%%%%%%%%%%%%%%%%

For the calculation of cross sections and the simulation of signal and
background events the generic event generator {\it SHERPA} was used
\cite{Gleisberg:2003xi}. This program is based on the matrix-element generator
{\it AMEGIC} \cite{Krauss:2001iv} and allows to simulate processes with up to
six particles in the final state. {\it SHERPA} also supports Supersymmetry and
uses {\it ISAJET~7.67} \cite{Paige:2003mg} for the generation of the mSUGRA
particle spectrum. The photon spectrum is taken into account by using the {\it
  CompAZ} parameterisation \cite{Zarnecki:2002qr}, which is well suited for
this analysis. 

The response of the detector has been simulated with {\it SIMDET}
\cite{Pohl:2002vk}, a parametric Monte Carlo for the TESLA $e^+e^-$ detector.
It includes tracking and calorimeter simulation and particle reconstruction.
An acceptance gap of the photon collider detector for polar angles below
$7^\circ$ is taken into account in the event reconstruction as the only
difference to the $e^+e^-$ detector \cite{Monig:2004jg}. \ 
\begin{figure}[htb]
\begin{centering}
\begin{tabular}{ccc}
\includegraphics[width=0.4\linewidth]{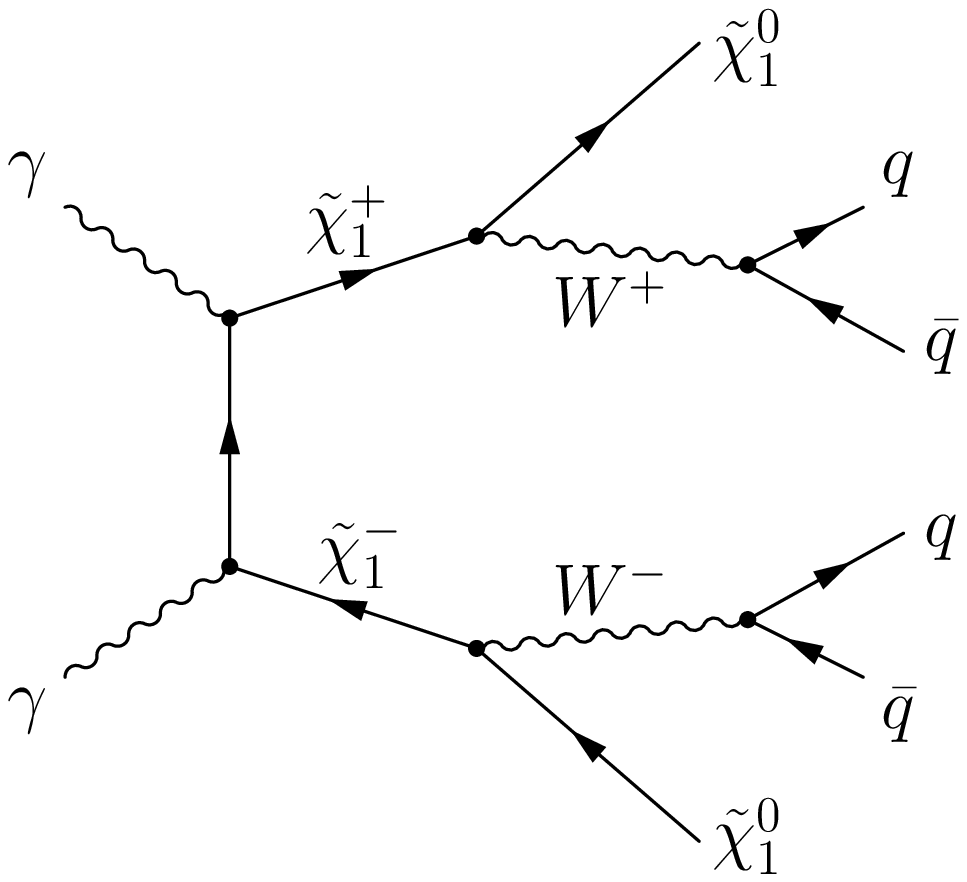}& \hspace{5mm} &
\includegraphics[width=0.38\linewidth]{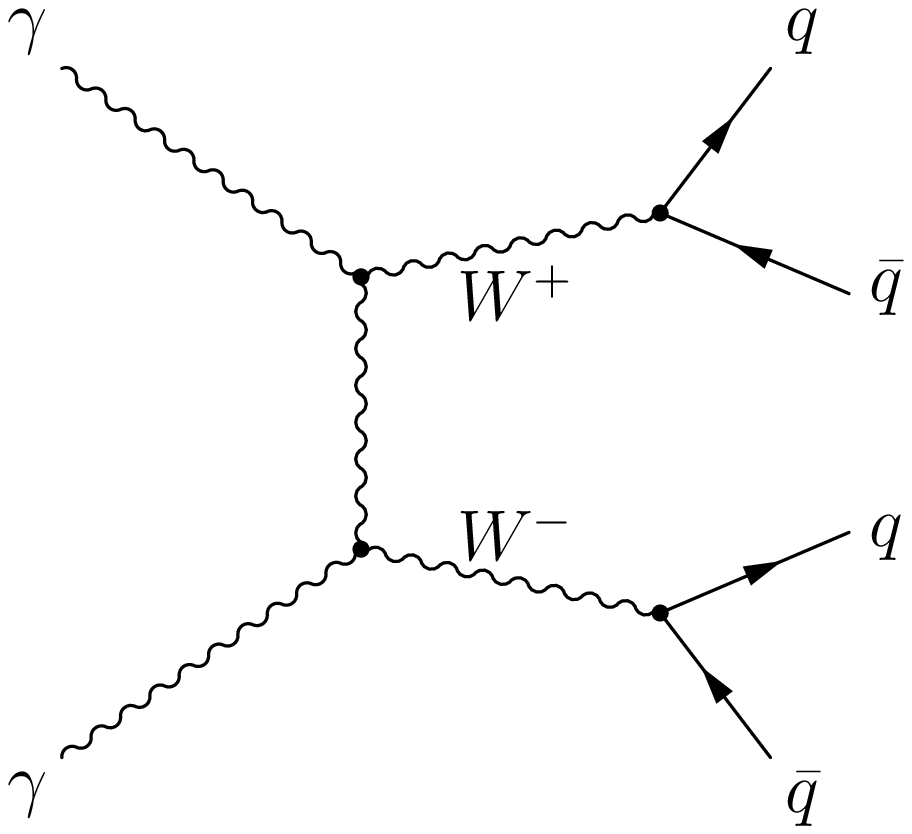}\\
a) & & b) \\
\includegraphics[width=0.38\linewidth]{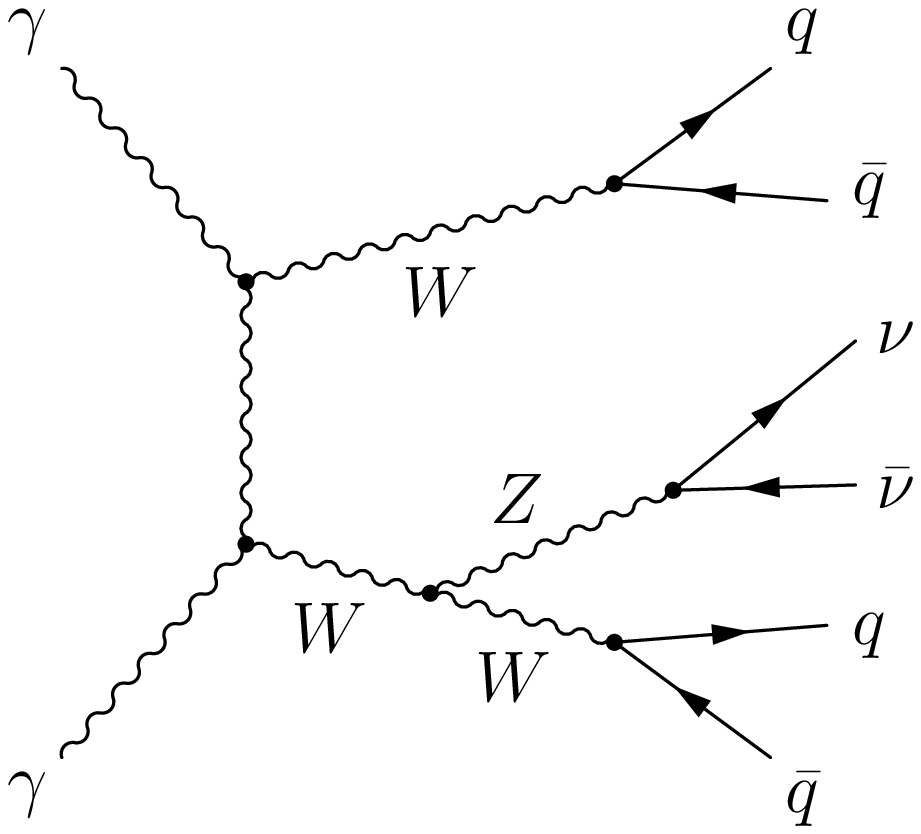}& \hspace{5mm} &
\includegraphics[width=0.4\linewidth]{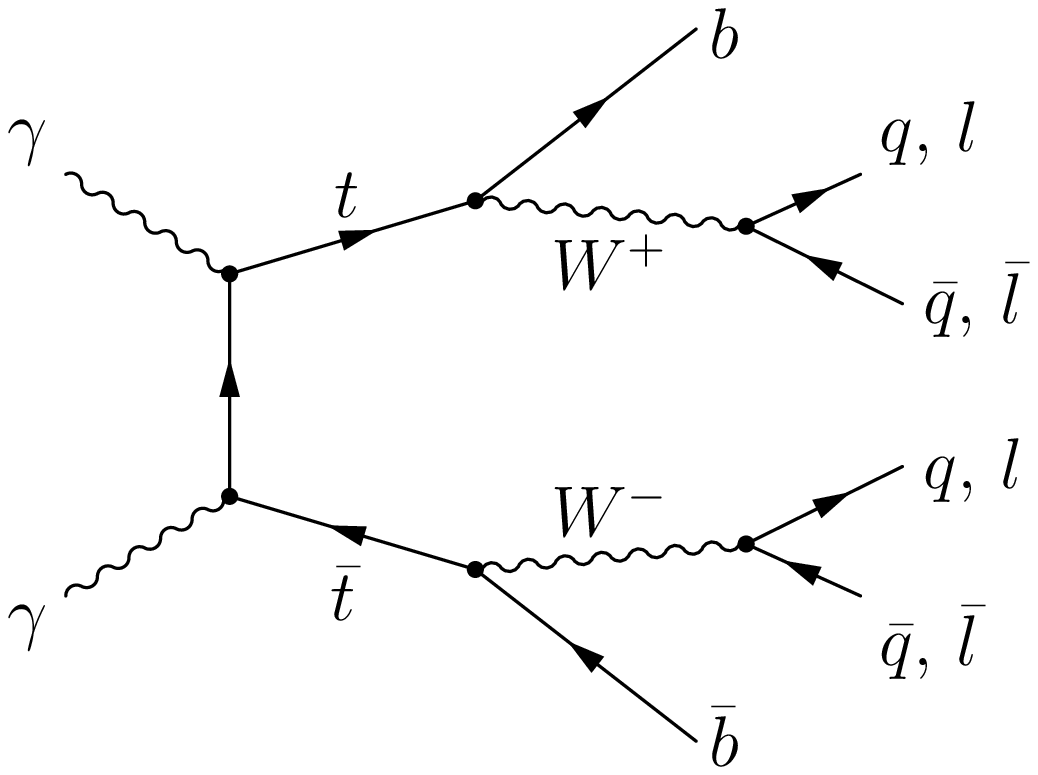}\\
c) & & d) \\
\end{tabular}
\end{centering}
\caption{Feynman diagrams for the signal process 
  $\gamma\gamma \rightarrow \chi^+_1\chi^-_1 \rightarrow
  q\bar{q}q\bar{q}\chi^0_1\chi^0_1$ (a) and for the background processes
  $\gamma\gamma \rightarrow 4$ jets (b), $\gamma\gamma \rightarrow
  W^+W^-Z^0 \rightarrow q\bar{q}q\bar{q}\nu\bar{\nu}$ (c), $\gamma\gamma
  \rightarrow t\bar{t} \rightarrow W^+W^-b\bar{b}$ (d).}
\label{fig:graphs}
\end{figure}
The signal is given by the process $\gamma\gamma \rightarrow
\tilde{\chi}_1^+\tilde{\chi}_1^-\rightarrow
W^+W^-\tilde{\chi}_1^0\tilde{\chi}_1^0 \rightarrow
q\bar{q}q\bar{q}\tilde{\chi}_1^0\tilde{\chi}_1^0$ (Fig.~\ref{fig:graphs}a),
where both charginos decay into a neutralino and a W-boson with a branching
ratio of $BR({\tilde{\chi}_1^\pm}\rightarrow {\tilde{\chi}_1^0}
W^\pm)=26.2\%$. The W-bosons are identified via their decay into hadrons
$BR(W^\pm\rightarrow q\bar{q})=68\%$. In the model used here, the neutralino
is the lightest supersymmetric particle (LSP) and stable. It cannot be
detected and therefore the signature for the signal is given by 4 jets plus
missing transverse momentum. The signal cross section is approximately given
by
\begin{equation}
\sigma_{sig} \approx
\sigma_p\cdot BR(\tilde{\chi}_1^\pm \rightarrow W^\pm \tilde{\chi}_1^0)^2 
\cdot BR(W^\pm\rightarrow q\bar{q})^2
\label{eqn:sigxs}
\end{equation}
in which $W$-bosons are assumed to be on-shell. However with {\it SHERPA} the
full process $\gamma\gamma \rightarrow
q\bar{q}q\bar{q}\tilde{\chi}_1^0\tilde{\chi}_1^0$ having 6 final state
particles was calculated, involving off-shell $W$-bosons. The diagram in
Fig.~\ref{fig:graphs}a yields the by far dominant contribution. The cross
sections are $\sigma_{sig}=2.62$\,fb for an electron centre-of-mass energy of
${\sqrt{s_{ee}}=500 \gev}$ and $\sigma_{sig}=7.98$\,fb for ${\sqrt{s_{ee}}=600
  \gev}$ (Table~\ref{tbl:allxs}). This corresponds to 2620 respectively 7980
signal events for an integrated luminosity of $1000fb^{-1}$ (one year).
The full 6-particle cross section is about 25\% larger than the simple
estimate using eqn.~\ref{eqn:sigxs} and the on-shell cross section and
branching ratios.  This comes roughly half from non double-resonant production
processes and from the fact that the phase space for the $\tilde{\chi}_1^0 W$
decay gets slightly larger with off-shell Ws. The non double-resonant
production processes are partially suppressed by the cut on the W-mass
explained later.

The major background is the Standard Model process $\gamma\gamma \rightarrow
4$ jets, for which Fig.~\ref{fig:graphs}b shows the main contribution via
$W$-pair production. Again the full 4 particle final state was simulated,
though only the light quarks $u, d, c, s$ and gluons were included. If the
electroweak subprocess is $\gamma \gamma \rightarrow q\bar{q}$ and the other
two jets stem from gluon radiation, the following parton shower is matched to
the 2nd order QCD matrix element to avoid double counting
\cite{Schalicke:2005nv}.  The top and bottom quarks were neglected, their
influence would be at the percent to per mille level. The calculated cross
sections for this background are 13.7 pb for ${\sqrt{s_{ee}}=500 \gev}$ and
13.4 pb for ${\sqrt{s_{ee}}=600 \gev}$ (Table~\ref{tbl:allxs}), which
corresponds to 13.7 (13.4) million events per year. Compared to the signal,
this is a difference of 3 to 4 orders of magnitude.

Two minor background sources have also been included: The process
$\gamma\gamma \rightarrow W^+W^-Z^0 \rightarrow q\bar{q}q\bar{q}\nu\bar{\nu}$
of $WWZ$ production (Fig.~\ref{fig:graphs}c), where the $W$-bosons decay to
hadrons and the $Z$-boson to undetectable neutrinos ($\nu_e$, $\nu_\mu$,
$\nu_\tau$). The second one is the production of top quarks that decay into a
$W^\pm$ and a $b$-quark $\gamma\gamma \rightarrow t\bar{t} \rightarrow
W^+W^-b\bar{b}$ (Fig.~\ref{fig:graphs}d). Here the decay of $W$-bosons into
leptons was also taken into account, because due to the $b$-quarks, a 4 jet
final state can occur even if one $W^\pm$ does not decay into quarks. These
two backgrounds have been simulated by generating $WWZ$ and $W^+W^-b\bar{b}$
events with {\it SHERPA} while doing the treatment of the decay with {\it
  PYTHIA} \cite{Sjostrand:2003wg}. The resulting cross sections that include
the decay branching ratios are summarised in Table~\ref{tbl:allxs}.

\begin{table}[hbt]
\renewcommand{\arraystretch}{1.2}
\centering
\begin{tabular}{|l||c|c|}
\hline
Channel & ${\sqrt{s_{ee}}=500 \gev}$ & ${\sqrt{s_{ee}}=600 \gev}$\\
\hline\hline
$\gamma\gamma \rightarrow \chi^+_1\chi^-_1 \rightarrow q\bar{q}q\bar{q}\chi^0_1\chi^0_1$ & 2.62 fb & 7.98 fb\\
$\gamma\gamma \rightarrow 4$ jets & 13.704 pb & 13.416 pb\\
$\gamma\gamma \rightarrow W^+W^-Z^0 \rightarrow q\bar{q}q\bar{q}\nu\bar{\nu}$ & 1.565 fb & 4.241 fb\\
$\gamma\gamma \rightarrow t\bar{t} \rightarrow W^+W^-b\bar{b}$ & 68.8 fb & 159.06 fb\\
\hline
\end{tabular}
\caption{Cross sections for the signal and background processes for the two 
  considered collision energies ${\sqrt{s_{ee}}=500 \gev}$ and 
  ${\sqrt{s_{ee}}=600 \gev}$.}
\label{tbl:allxs}
\end{table}

There is another, inherent source of background of low energetic hadrons. For
the considered energies, the cross-section for $\gamma\gamma \rightarrow
q\bar{q}$ events is several hundreds of nb so that on average 1.8 such events are
produced per bunch crossing (pileup) that overlay the high energy events
\cite{schulte}. The
pileup events were produced with {\it PYTHIA}, while the overlay is done
within {\it SIMDET}.

%%%%%%%%%%%%%%%%%%%%%%%%%%%%%%%%%%%%%%%%%%%%%%%%%%%%%%%%%%%%%%%%%%%%%%%
\section{Event analysis}
%%%%%%%%%%%%%%%%%%%%%%%%%%%%%%%%%%%%%%%%%%%%%%%%%%%%%%%%%%%%%%%%%%%%%%%

The first step in the event analysis is to reject pileup tracks as much as
possible, in order to reduce their contribution to the high energy signal
tracks.  For this purpose, the measurement of the impact parameter of a
particle along the beam axis with respect to the primary vertex is used.

\begin{figure}[bht]
\begin{tabular}{cc}
\includegraphics[width=0.48\linewidth]{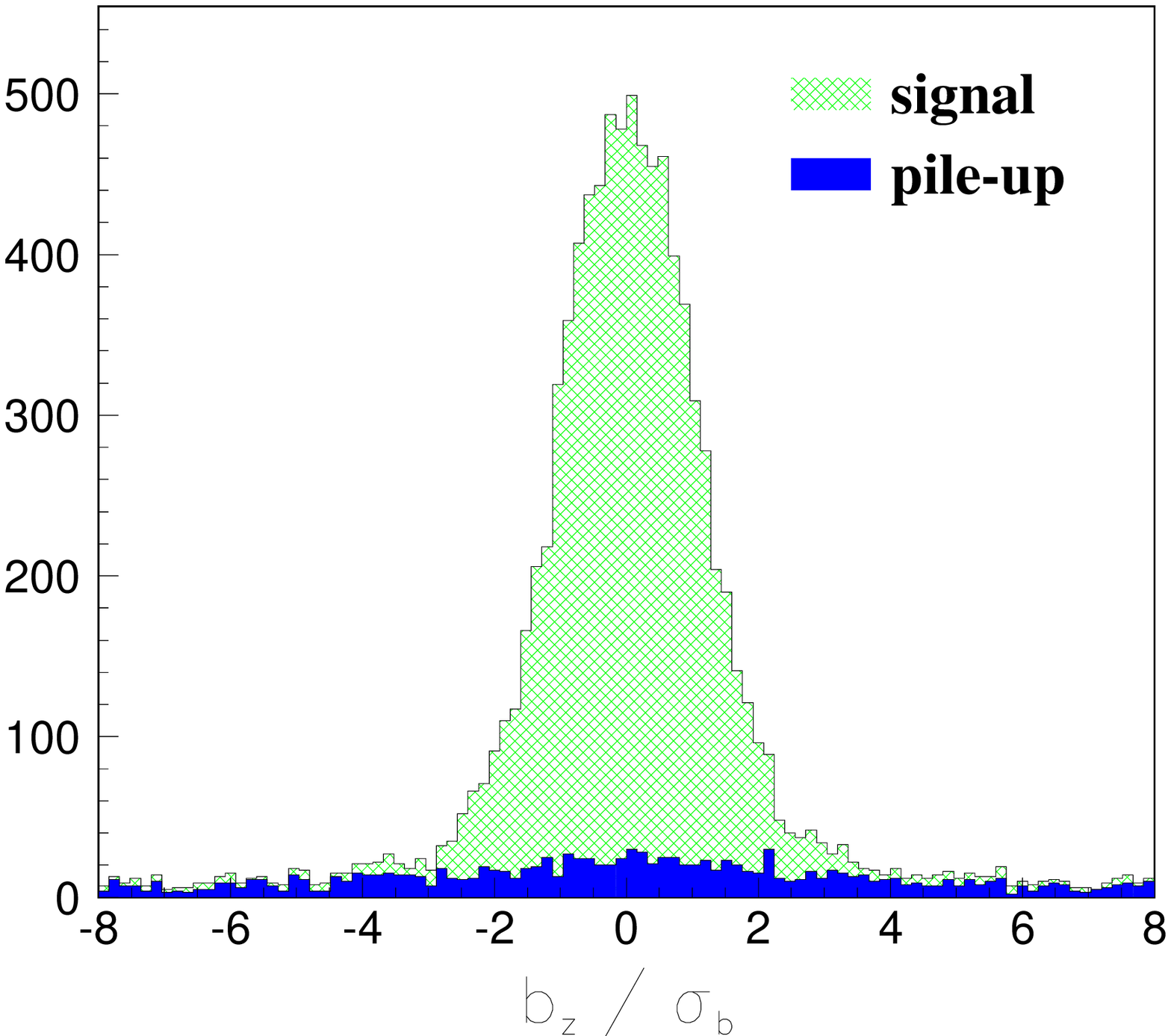}&
\includegraphics[width=0.48\linewidth]{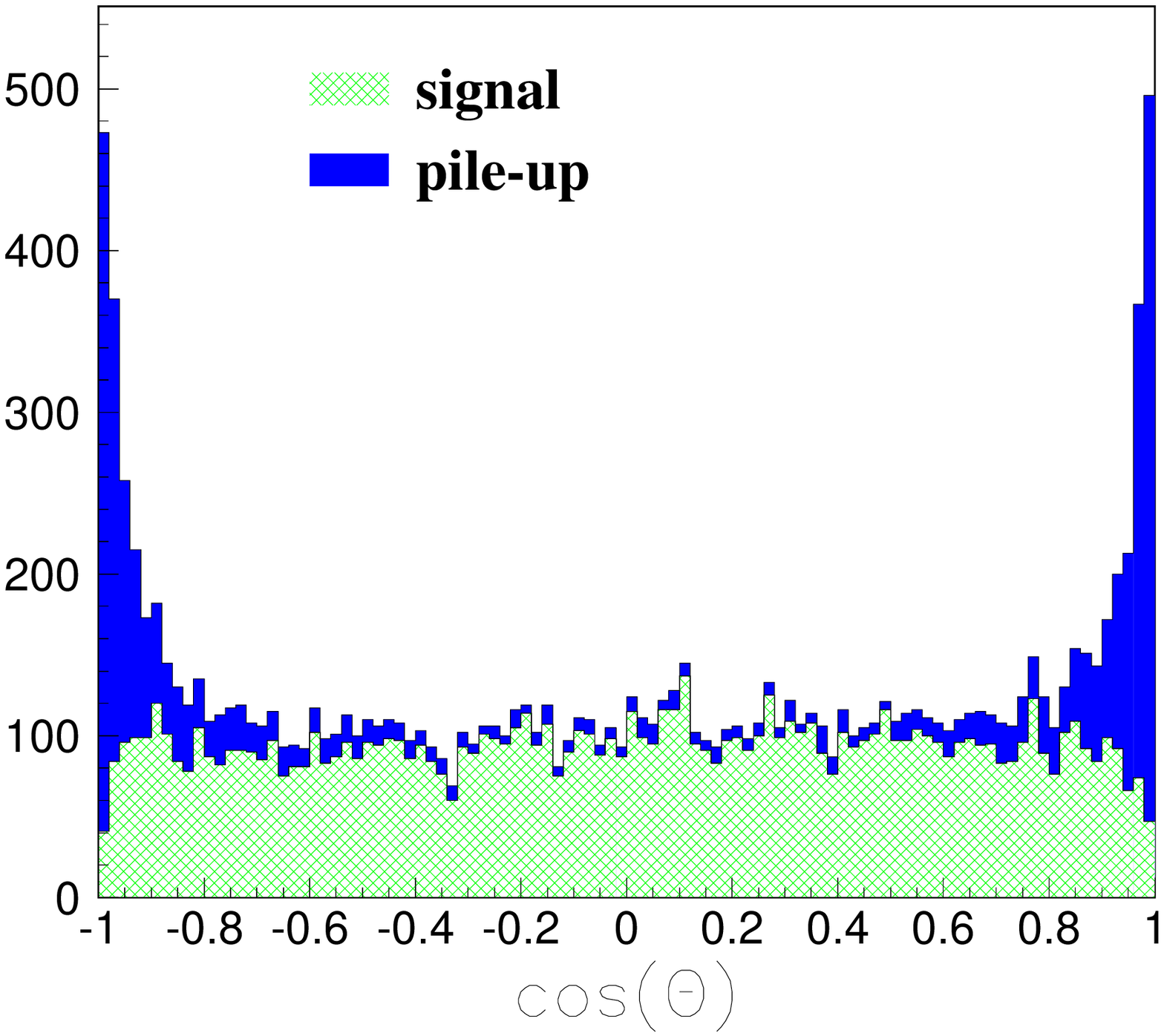}\\
\end{tabular}
\caption{Left: The distribution of the impact parameter $b_z$ with respect 
to the primary vertex divided by its measurement error $\sigma_{b_z}$ for 
signal and pileup tracks. Right: The distribution of the cosine of the polar 
angle $\theta$ for signal and pileup tracks.}
\label{fig:tracks}
\end{figure}
The beamspot length for TESLA is about $300 \mu m$, while the measurement
error for the impact parameter is only $\approx 5 \mu m$.  Using the precise
measurements from the vertex detector, the primary vertex is first
reconstructed as the momentum weighted average z-impact parameter\footnote{The
  z-impact parameter is defined as the $z$ coordinate of the impact point in
  the $x-y$ plane.} of all tracks in the event.

The difference $b_z$ of the z-impact parameter with respect to the primary
vertex, divided by the measurement error $\sigma_{b_z}$ is shown in
Fig.~\ref{fig:tracks} (left) for signal and pileup tracks. Since the
distribution for the pileup tracks is much broader than for the signal, only
tracks with $|b_z|<3\cdot\sigma_{b_z}$ are accepted for further event
analysis.

The polar angle of each track, i.e. the angle with the beam axis is a further
possibility to reduce the pileup. Because of the $t-$channel production
mechanism, the pileup tracks are concentrated at low polar angles 
(Fig.~\ref{fig:tracks}, right).
Only tracks with a polar angle larger than $18^\circ$
(i.e. $|\cos\theta|<0.95$) are kept.

For the reconstruction of jets the standard {\it PYTHIA} cluster finding
algorithm is used\footnote{The minimum distance parameter was set to
  $d_{join}=6.3 \gev$.}, with the constraint of at least 4 reconstructed jets.
The jets are sorted by their transverse momentum $p_T$. 
The low $p_T$ jets are very much dominated by pileup tracks, therefore only the 4 jets with
the highest $p_T$ are taken for the reconstruction of the two
$W$-bosons. This
is done by combining\footnote{The combinatorics are such that the $W_1$
  always contains the jet with highest $p_T$.} pairs of jets in such a way
that the invariant 2-jet masses $m(W_1), m(W_2)$, i.e. the reconstructed
$W$-masses deviate minimally from the on-shell $W$-mass $m_W=80.4 \gev$.

%\begin{equation}
%[m(W_1)-m_W]^2+[m(W_2)-m_W]^2\rightarrow \mbox{min!}
%\end{equation}
\begin{figure}
\begin{tabular}{cc}
\includegraphics[width=0.48\linewidth]{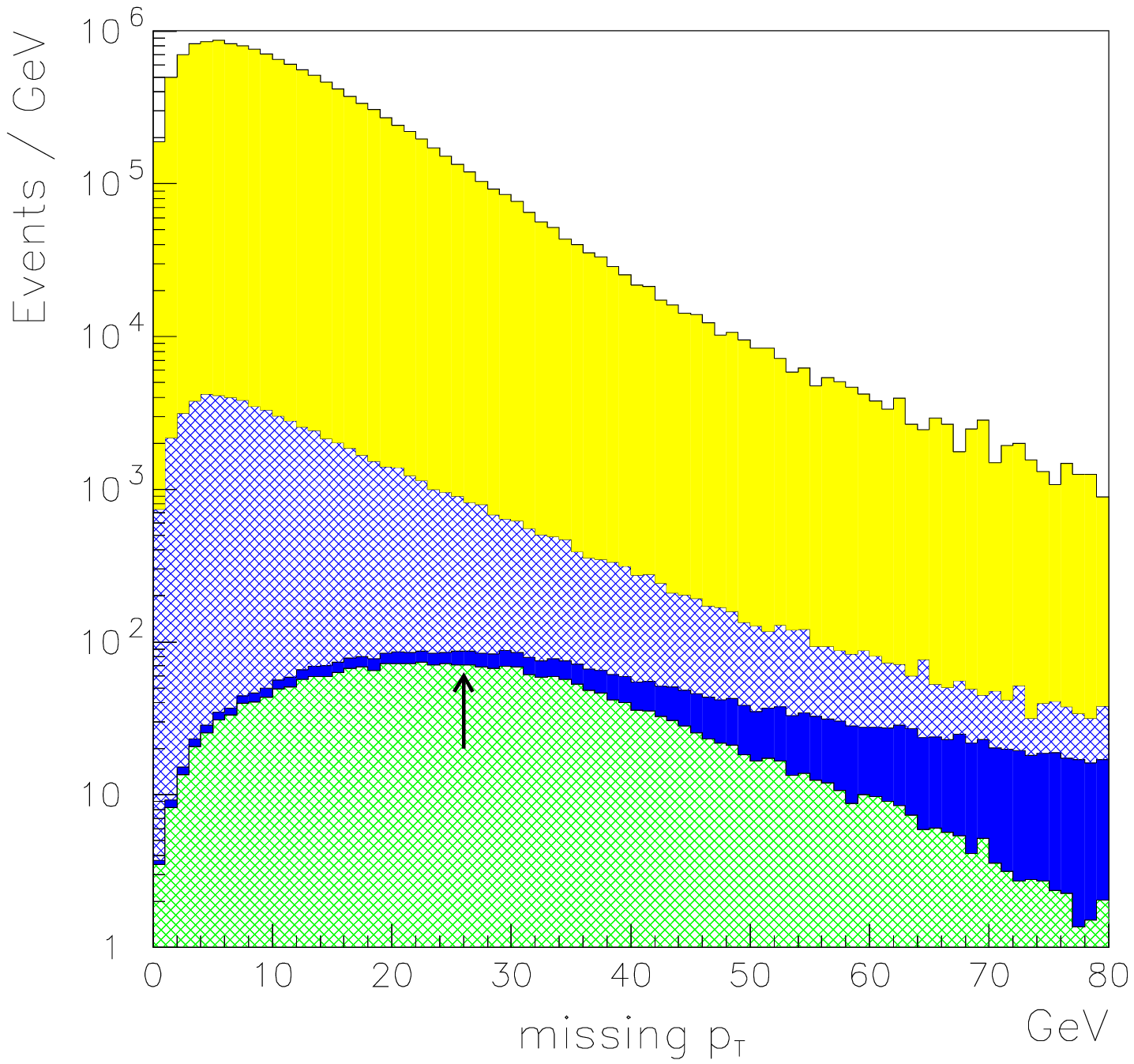}&
\includegraphics[width=0.48\linewidth]{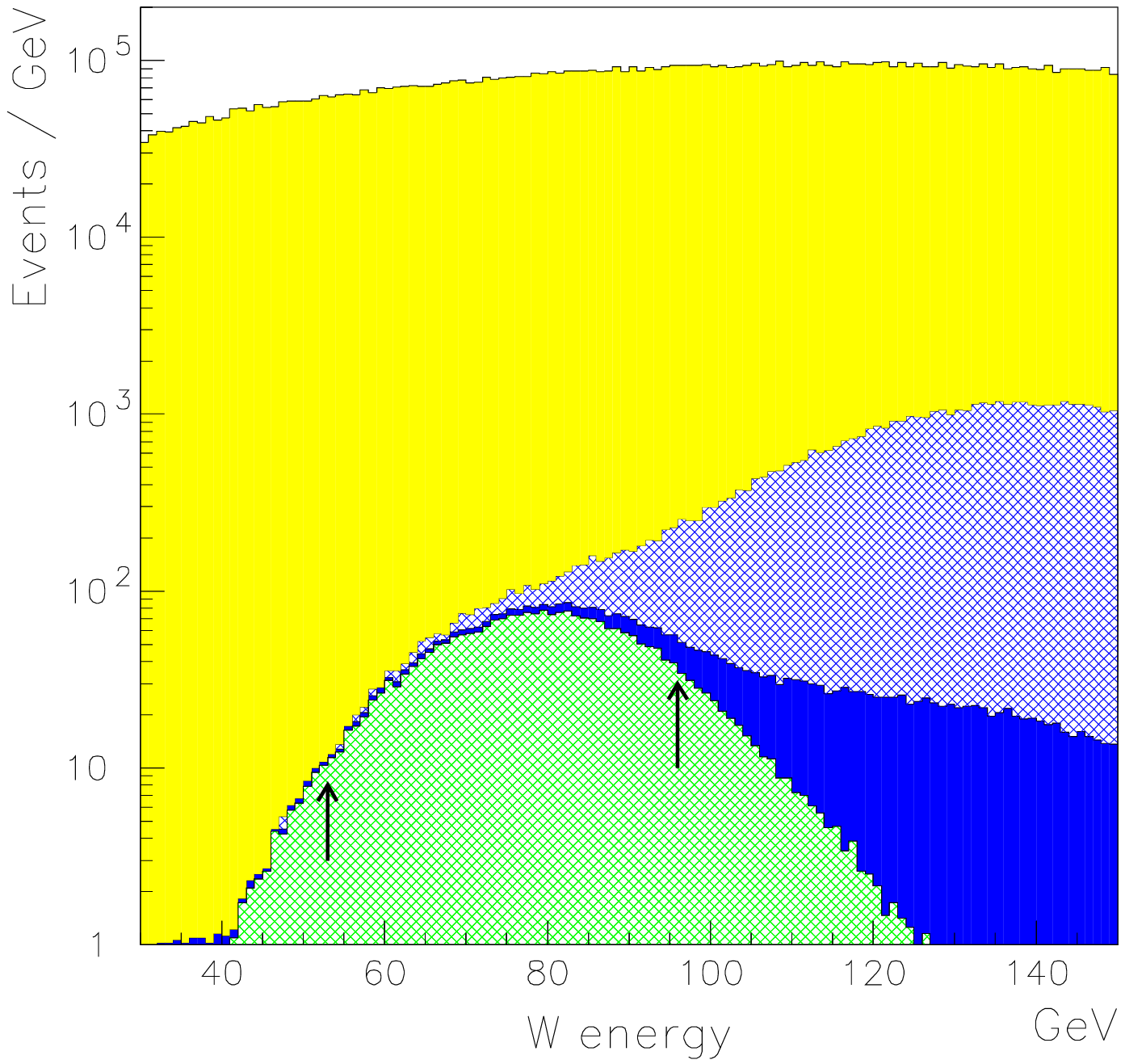}\\
a) & b)\\
\includegraphics[width=0.48\linewidth]{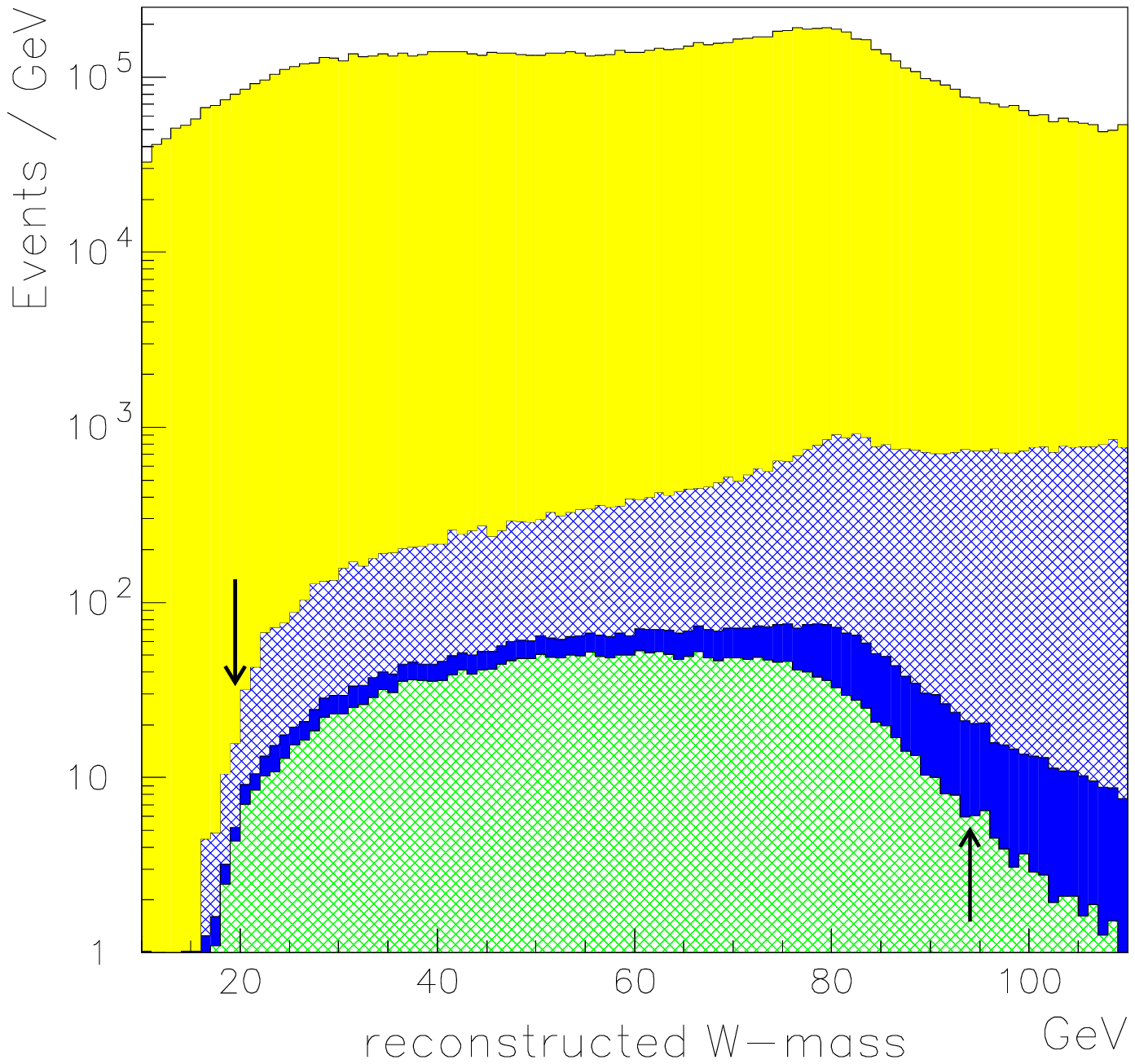}&
\includegraphics[width=0.48\linewidth]{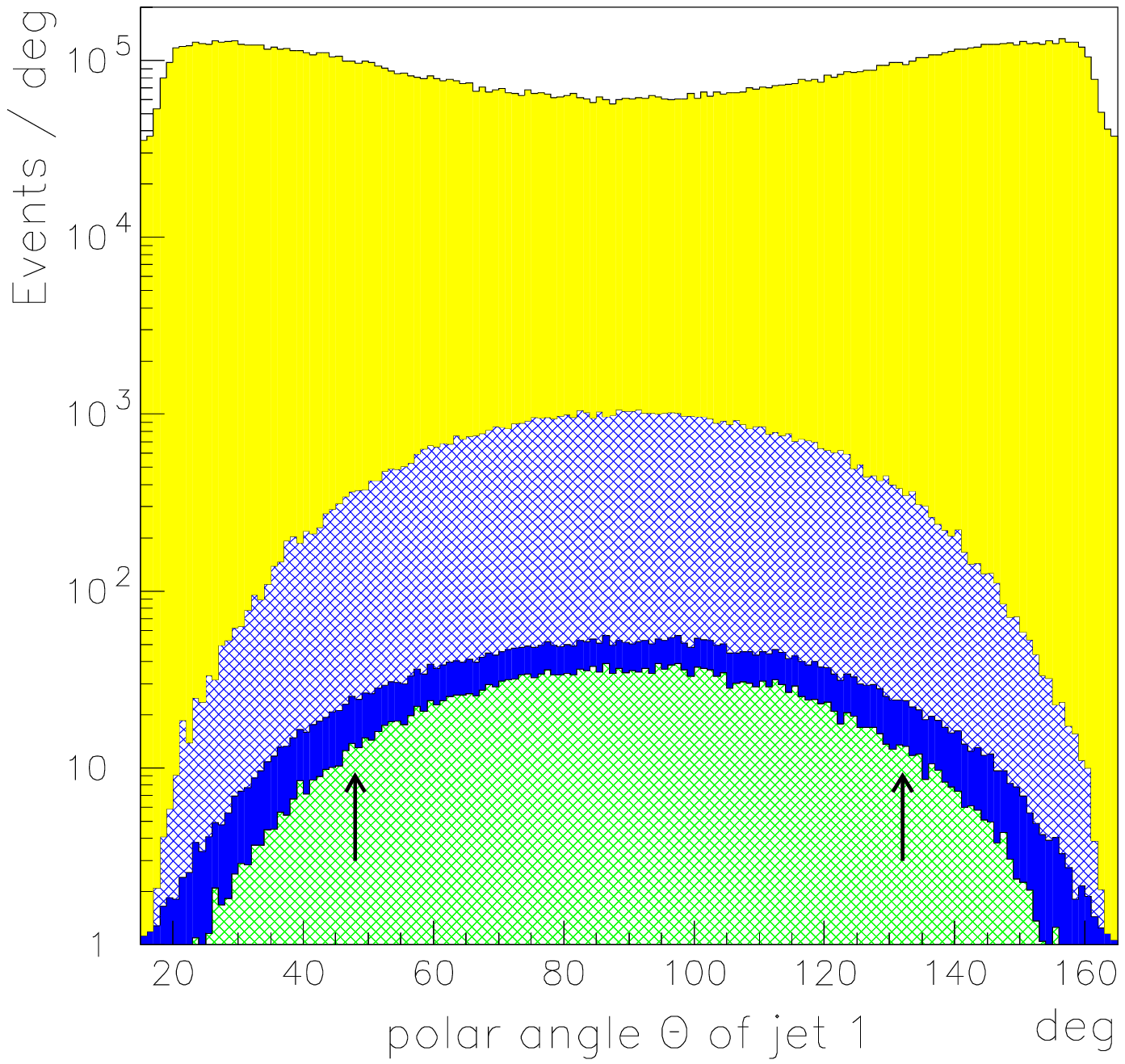}\\
c) & d)\\
\end{tabular}
\caption{For $\sqrt{s_{ee}}=500 \gev$: a) The missing $p_T$ distribution. b)
  The energy distribution of the reconstructed $W$-boson $W_1$. c) The
  invariant mass of the reconstructed $W_1$. d) The polar angle of the jet
  with highest $p_T$. The arrows indicate the applied cuts. The green hatched
  (light hatched) area represents the signal. The blue (dark) area are the
  $\gamma\gamma \rightarrow W^+W^-Z^0$ events. The blue hatched (dark hatched)
  contribution corresponds to $\gamma\gamma \rightarrow t\bar{t}$ events,
  while the yellow (light) area represents the $\gamma\gamma \rightarrow
  4$ jets events. }

\label{fig:ptener}
\end{figure}
\begin{table}[bht]
\begin{tabular}{|l||c|c||c|c|}
\hline
\multirow{2}{2cm}{Observable} & \multicolumn{2}{c||}{$\sqrt{s_{ee}}=500 \gev$} & \multicolumn{2}{c|}{$\sqrt{s_{ee}}=600 \gev$} \\
\cline{2-5}
 & min. &max. & min. & max.\\
\hline
\hline
acoplanarity & 0.225 rad & $\pi$ & 0.09 rad & $\pi$ \\
missing $p_T$ & 26 \gev& $-$& 22 \gev& $-$ \\
thrust & $-$ & 0.973& $-$ & 0.983 \\
\hline
energy of $W_1$ & 53 \gev& 96 \gev& 65 \gev& 122 \gev \\
energy of $W_2$ & 50 \gev& 99 \gev& 58 \gev& 124 \gev\\
lepton - energy & $-$& 14 \gev& $-$& 20\gev \\
total energy & 132 \gev& 226 \gev & 110 \gev& 262 \gev\\
reconstructed $W$ - mass & 19.5 \gev& 94 \gev& 23 \gev& 116 \gev\\
visible mass & 108 \gev& 235 \gev& 100 \gev& 280 \gev \\
\hline
polar angle of 1st jet & 0.84 rad & 2.30 rad&0.82 rad & 2.32 rad\\
polar angle of 2nd jet & 0.63 rad&2.51 rad &0.58 rad & 2.56 rad\\
polar angle of 3rd jet & 0.4 rad&2.74 rad &0.44 rad &2.70 rad \\
polar angle of 4th jet & 0.3 rad&2.84 rad &0.32 rad &2.82 rad \\
larger polar angle of $W$s & 1.35 rad & $\pi$& 1.35 rad & $\pi$\\
smaller polar angle of $W$s & $-$ & 1.8 rad& $-$ & 1.85 rad\\
\hline
\end{tabular}
\caption{The cut variables that are used in the event analysis for
  $\sqrt{s_{ee}}=500 \gev$ and $\sqrt{s_{ee}}=600 \gev$. The $min.$/$max.$
  values define the range in which the variables have to be so that an event
  is accepted.}
\label{tbl:cutvars}
\end{table}
In order to improve the signal to background ratio, cuts were applied on
various calculated observables. Table~\ref{tbl:cutvars} lists all considered
variables together with the applied cut condition for the $\sqrt{s_{ee}}=500
\gev$ and $\sqrt{s_{ee}}=600 \gev$ case. Only events that fulfil all cut
conditions are accepted and considered as signal-like. The cuts have
been optimised by varying the cut conditions one after another and 
fixing them to the values with best resulting statistical error.

\begin{figure}[hbt]
\begin{tabular}{cc}
\includegraphics[width=0.48\linewidth]{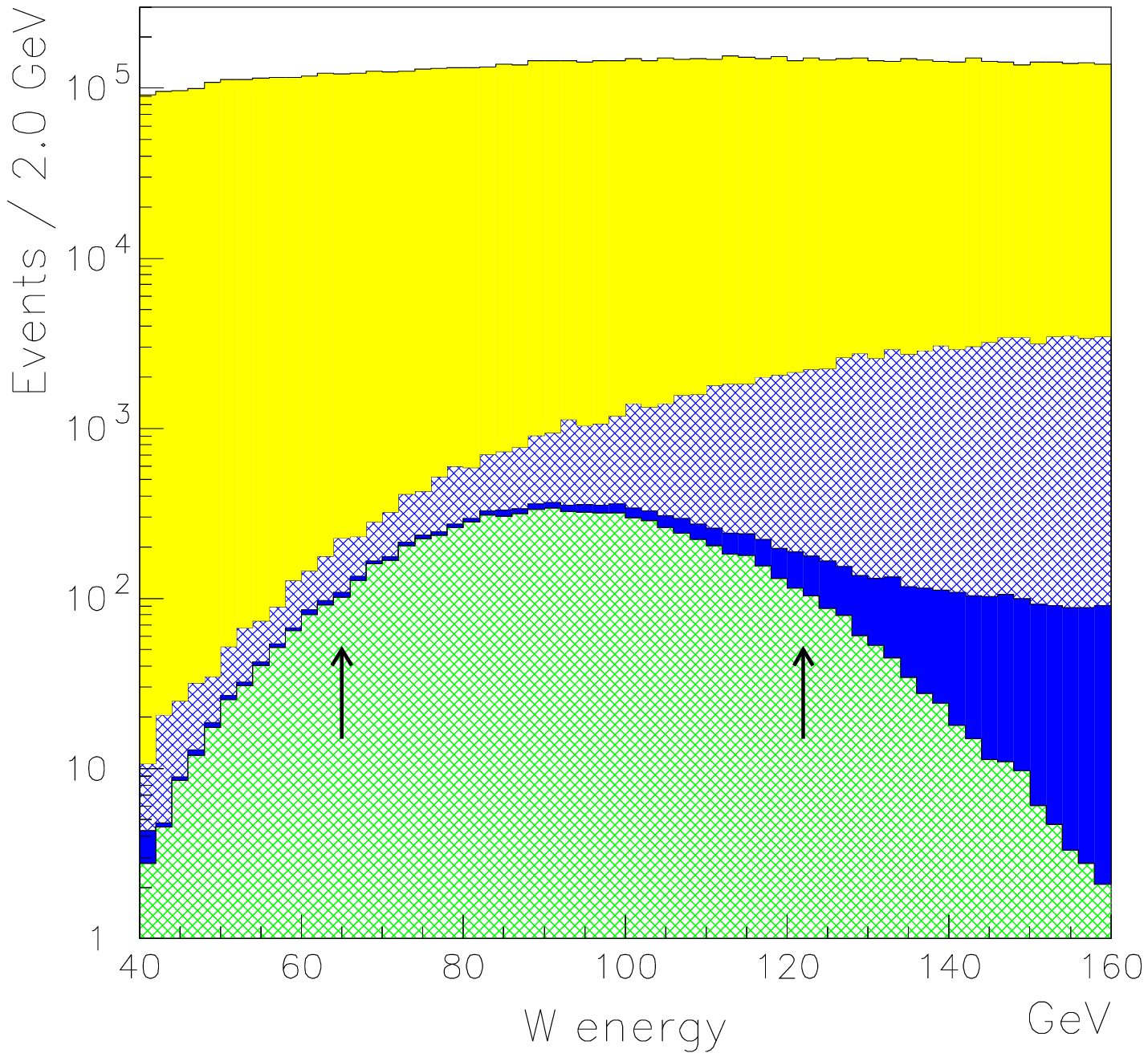}&
\includegraphics[width=0.48\linewidth]{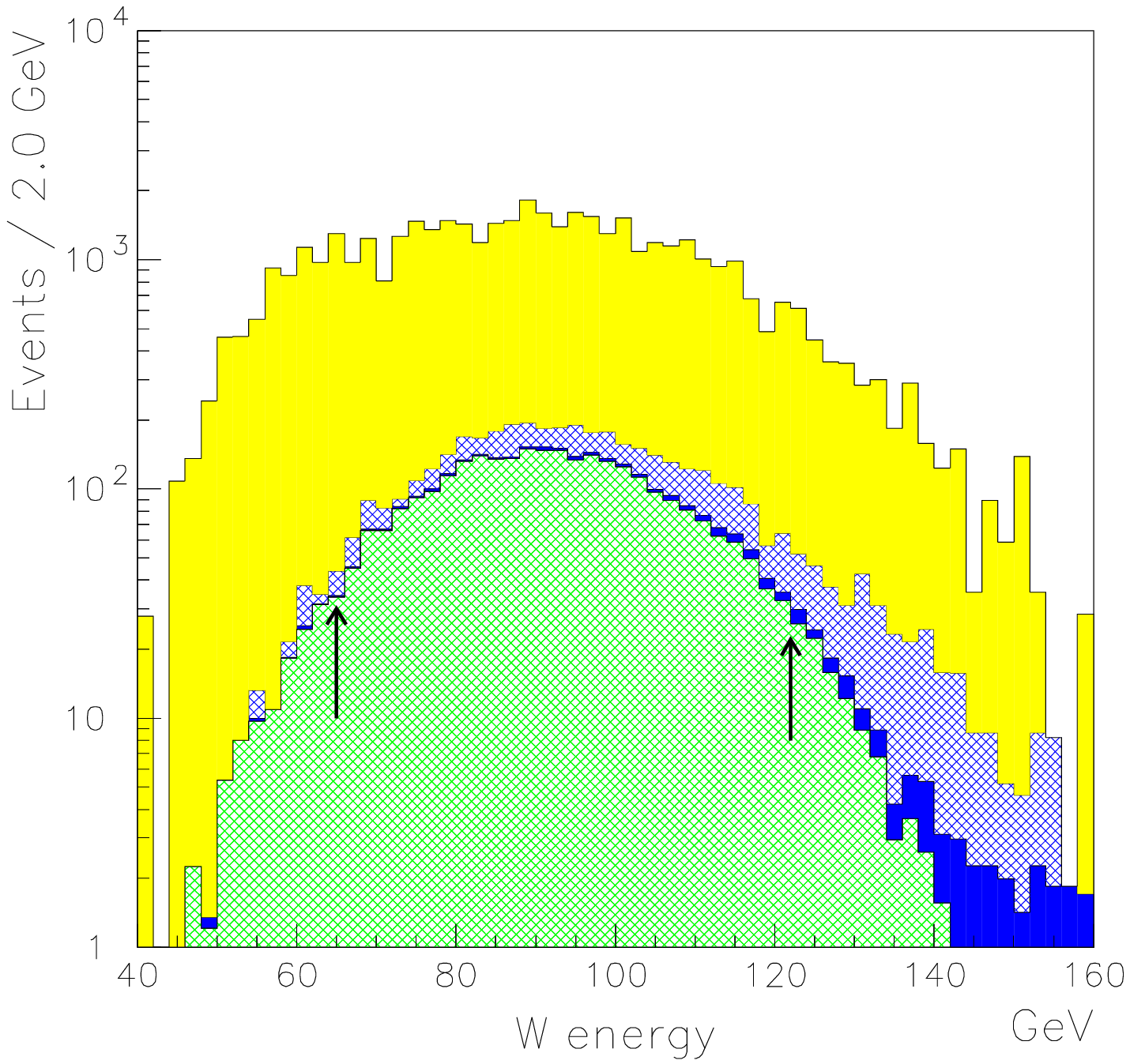}\\
\end{tabular}
\caption{Left: The energy distribution of the reconstructed $W$-boson $W_1$
  for $\sqrt{s_{ee}}=600 \gev$. Right: The same distribution after 
applying all cuts except the one on the $W_1$-energy. The arrows 
indicate the cut conditions. The green hatched (light hatched) 
corresponds to signal, blue (dark) to $\gamma\gamma \rightarrow
  W^+W^-Z^0$, blue hatched (dark hatched) to $\gamma\gamma \rightarrow
  t\bar{t}$ and yellow (light) to the $\gamma\gamma \rightarrow
  4$ jets events.}
\label{fig:cut}
\end{figure}

The acoplanarity is defined as $\pi-\delta$, where $\delta$ is the angle
between the two reconstructed $W$-bosons in the x-y plane. The distribution of
the missing transverse momentum is shown in Fig.~\ref{fig:ptener}a for the
signal and the three considered backgrounds for $\sqrt{s_{ee}}=500 \gev$. The
logarithmic scale illustrates the huge amount of background compared to the
signal. Fig.~\ref{fig:ptener}b and \ref{fig:ptener}c show distribution of
energy and reconstructed mass of $W_1$. 
The cut on the reconstructed W-mass comes out fairly asymmetric around the
nominal W-mass because the phase space of the chargino decay favours low mass
W-bosons and in addition the usage of only four jets in the analysis, which is
needed to reject pileup tracks, biases the reconstruction towards low masses.
Further cut variables are the polar
angles of the 4 jets that were used for the $W$
reconstruction. Fig.~\ref{fig:ptener}d shows the distribution for the jet with
highest $p_T$. 
The applied
cuts strongly improve the signal to background ratio. Fig.~\ref{fig:cut}
illustrates this for the $\sqrt{s_{ee}}=600 \gev$ case. It shows the energy
distribution of a reconstructed $W$ before and after cuts were applied.

Table~\ref{tbl:cuts} summarises the cut efficiency, showing the 
number of events for the signal and the background channels for an 
integrated luminosity of $1000fb^{-1}$ before and after cuts.\\

\begin{table}[hbt]
\centering
\renewcommand{\arraystretch}{1.2}
\begin{tabular}{|l||c|c|c|c|}
\hline
 & signal & 4 jets & $W^+W^-Z^0$ & $t\bar{t}$\\
\hline\hline
{${\sqrt{s_{ee}}=500}$ GeV} & & & &\\
 without cuts & $2620$ & $13.7\cdot10^6$ & $1565$ & $68.8\cdot10^3$\\
 after cuts & 453 & 4065 & 15 & 4 \\
\hline
{${\sqrt{s_{ee}}=600}$ GeV} & & & &\\
 without cuts & $7976$ & $13.4\cdot10^6$ & 4241 &$159.1\cdot10^3$\\
 after cuts & $1925$ & $14760$ & 81 & 776\\
\hline
\end{tabular}
\caption{Number of events per year ($1000fb^{-1}$) for signal and background 
  channels before and after cuts. }
\label{tbl:cuts}
\end{table}
%%%%%%%%%%%%%%%%%%%%%%%%%%%%%%%%%%%%%%%%%%%%%%%%%%%%%%%%%%%%%%%%%%%%%%%
\section{Results}
%%%%%%%%%%%%%%%%%%%%%%%%%%%%%%%%%%%%%%%%%%%%%%%%%%%%%%%%%%%%%%%%%%%%%%%

An efficiency of 17.3\% and a purity of 10.0\% was obtained for an electron
beam centre-of-mass energy of ${\sqrt{s_{ee}}=500 \gev}$, resulting in a
statistical error of 14.9\% (Table~\ref{tbl:errors})\footnote{${\Delta
    N}/{N}={1}/{\sqrt{\varepsilon\cdot p\cdot N}}$, where $\varepsilon$ is the
  efficiency, $p$ the purity and $N$ the total number of signal events.}. For
${\sqrt{s_{ee}}=600 \gev}$ an efficiency of 24.1\% and a purity of 11.0\% was
obtained, resulting in a statistical error of 6.9\%. Because of the higher
signal cross section, the statistical error gets smaller for $600 \gev$
compared to $500 \gev$. However, generally the final errors are quite large.
This has a couple of reasons: The Standard Model background $\gamma\gamma
\rightarrow 4$ jets has a cross section very much larger than the
signal.  The distinction of signal and background events is more difficult in
comparison with the $e^+e^-$-collider.  There is no fixed beam energy that
could be used for kinematic constraints (on the $W$-energy for instance). In
addition, particles with polar angles below $7^\circ$ are not detected, which
makes the $p_T$ and acoplanarity cuts less effective.

\begin{table}[hbt]
\centering
\renewcommand{\arraystretch}{1.2}
\begin{tabular}{|l||c|c|c|c|c|}
\hline
 $\mathbf{\sqrt{s_{ee}}}$ & \multirow{2}{2.4cm}{signal events per year} &
 \multirow{2}{2.8cm}{background events per year}&
 \multirow{2}{1.5cm}{\centering efficiency\\ $\varepsilon$}
 &\multirow{2}{1.3cm}{\centering purity\\ 
 $p$}&  \multirow{2}{1.9cm}{stat. error $\Delta N/N$ }\\
 & & & & &\\
\hline\hline
$500 \GeV$ & $453$ & $4084$ & $17.3\%$ & $10.0\%$ & \bf 14.9\%\\ 
$600 \GeV$ & $1925$ & $15.6\cdot10^3$ & $24.1\%$ & $11.0\%$ &\bf 6.9\%\\
\hline
\end{tabular}
\caption{Number of signal events and the total number of background events
 after all cuts for $1000fb^{-1}$. In addition the final efficiencies,
 purities and statistical errors.}
\label{tbl:errors}
\end{table}

Using equation \ref{eqn:sigxs} the statistical error for the branching ratio
BR(${\tilde{\chi}_1^\pm}\rightarrow {\tilde{\chi}_1^0} W^\pm$) can be derived.
We neglect the error of the luminosity, which is supposed to be on the per
mille level. Since the chargino mass will be precisely measured at the Linear
Collider, the pair production cross section is known. Therefore, the relative
error for BR(${\tilde{\chi}_1^\pm}\rightarrow {\tilde{\chi}_1^0} W^\pm$) is
simply one half of the statistical error ${\Delta N}/{N}$, because
the branching ratio enters quadratically in the total cross section.

Thus the result of this analysis is an expected statistical error for the
directly measured branching ratio BR(${\tilde{\chi}_1^\pm}\rightarrow
{\tilde{\chi}_1^0} W^\pm$) of 7.5\% for ${\sqrt{s_{ee}}=500 \gev}$ and 3.5\%
for ${\sqrt{s_{ee}}=600 \gev}$.

%%%%%%%%%%%%%%%%%%%%%%%%%%%%%%%%%%%%%%%%%%%%%%%%%%%%%%%%%%%%%%%%%%%%%%%
\section{Interpretation with {\it Fittino}}
%%%%%%%%%%%%%%%%%%%%%%%%%%%%%%%%%%%%%%%%%%%%%%%%%%%%%%%%%%%%%%%%%%%%%%%

In \cite{Weiglein:2004hn} a global fit of the MSSM parameters for the 
SPS1a scenario has been presented, which was done with the program 
{\it Fittino} \cite{Bechtle:2004pc}. A set of 24 free parameters was 
fitted, based on a collection of simulated LHC and LC measurements with 
estimated uncertainties. 

We have repeated that fit for the scenario used in this analysis and included
the chargino branching ratio with its estimated measurement error as an
additional observable. For this purpose the low energy MSSM parameters and
observables that correspond to the mSUGRA parameters, which were selected for
this analysis, have been calculated with {\it SPHENO} \cite{Porod:2003um}
first. Table~\ref{tbl:obs} shows the list of all included observables.  The
estimated measurement errors were taken from \cite{Weiglein:2004hn} and scaled
according to the change in the measurement values with respect to those used
in the SPS1a fit. The numbers (e.g. the chargino mass
$m_{\tilde{\chi}_1^{\pm}}$) also differ slightly from the ones that were used
as input for the Monte Carlo analysis.  Those have been calculated with {\it
  ISAJET}, while {\it Fittino} uses {\it SPHENO} for the generation of the
SUSY particle spectrum.

However, only a subset of parameters has been fitted here for reasons of
simplicity. Table~\ref{tbl:fixed} shows the parameters that have been fixed
to their input values. They concern the squark sector, which is assumed not to
be very much influenced by a measurement of the chargino branching ratio.
\begin{table}[hbt]
\begin{center}
{
\begin{tabular}{|l|| c |c|}
\hline 
Measurement & Value & Uncertainty \\ 
\hline\hline
%$m_{\text{Z}}$       &     \text{91.1187 GeV} & \text{0.0021 GeV}\\ %
%$m_{\text{W}}$       &     \text{80.3422 GeV} & \text{0.039  GeV}\\ %
%$m_{\text{c}}$       &     \text{1.2     GeV} & \text{0.2    GeV}\\ %
%$m_{\text{b}}$       &     \text{4.2     GeV} & \text{0.5    GeV}\\ %
%$m_{\text{t}}$       &     \text{174.3   GeV} & \text{0.3    GeV}\\ %
%$m_{\tau}$           &     \text{1.77699   GeV} & \text{0.00029    GeV}\\ %
%$\alpha_s$           &     \text{0.1172}      & \text{0.0002}\\
%$G_F$                &     \text{1.16639$\cdot$10$^{\text{-5}}$ GeV$^{-2}$} & \text{1$\cdot$10$^{\text{-11}}$ GeV$^{-2}$}\\
%$1/\alpha$           &     \text{127.934}     & \text{0.027}\\
%$\sin^2 \theta_W$    &     \text{0.23113}     & \text{0.00015}\\
$m_{\text{h}^0}$     &     \text{110.6 GeV} & \text{0.5 GeV  }\\ %   # +- 3 GeV
$m_{\text{H}^0}$     &     \text{407.3 GeV} & \text{1.3 GeV   }\\ %   # not in LHC
$m_{\text{A}^0}$     &     \text{406.6 GeV} & \text{1.3 GeV   }\\ %   # not in LHC
$m_{\text{H}^{\pm}}$ &     \text{415.8 GeV} & \text{1.1 GeV   }\\ %   # not in LHC
%$m_{\tilde{\text{u}}_L}$ &     \text{561.0 GeV} & \text{9.4 GeV   }\\ %
%$m_{\tilde{\text{u}}_R}$ &     \text{545.2 GeV} & \text{23.7 GeV  }\\ %
%$m_{\tilde{\text{d}}_L}$ &  \text{565.6 GeV} & \text{9.5 GeV   }\\ %
%$m_{\tilde{\text{d}}_R}$ &  \text{545.0 GeV} & \text{22.7 GeV  }\\ %
%$m_{\tilde{\text{c}}_L}$ &     \text{561.0 GeV} & \text{9.4 GeV   }\\ %    # ~c_L
%$m_{\tilde{\text{c}}_R}$ &     \text{545.2 GeV} & \text{23.7 GeV  }\\ %    # ~c_R
%$m_{\tilde{\text{s}}_L}$ &     \text{565.6 GeV} & \text{9.4 GeV   }\\ %
%$m_{\tilde{\text{s}}_R}$ &     \text{545.0 GeV} & \text{23.7 GeV  }\\ %
%$m_{\tilde{\text{t}}_1}$ &     \text{399.7 GeV} & \text{1.9 GeV   }\\ %    # ~t_1 # not at LHC
%$m_{\tilde{\text{b}}_1}$ &     \text{514.0 GeV} & \text{5.5  GeV  }\\ %    # ~b_1
%$m_{\tilde{\text{b}}_2}$ &     \text{545.3 GeV} & \text{6.0  GeV  }\\ %    # ~b_2
$m_{\tilde{\nu}_{\text{e}L}}$ &     \text{209.2 GeV} & \text{0.8 GeV   }\\ %    # ~nu_eL        # not at LHC
$m_{\tilde{\text{e}}_L}$ &     \text{223.7 GeV} & \text{0.2 GeV   }\\ %    # ~e_L-
$m_{\tilde{\text{e}}_R}$ &     \text{166.2 GeV} & \text{0.06 GeV  }\\ %    # ~e_R-
$m_{\tilde{\mu}_L}$      &     \text{223.7 GeV} & \text{0.5 GeV   }\\ %    # ~mu_L-
$m_{\tilde{\mu}_R}$      &     \text{166.2 GeV} & \text{0.2 GeV   }\\ %    # ~mu_R-
$m_{\tilde{\tau}_1}$     &     \text{159.2  GeV} & \text{0.4 GeV   }\\ %    # ~tau_1-
$m_{\tilde{\tau}_2}$     &     \text{226.4 GeV} & \text{1.2 GeV   }\\ %    # ~tau_2- # not at LHC
$m_{\tilde{\text{g}}}$          &     \text{600.5  GeV} & \text{6.1 GeV   }\\ %    # ~g

$m_{\tilde{\chi}_1^0}$   &     \text{94.86 GeV} & \text{0.05 GeV  }\\ %    # ~chi_10
$m_{\tilde{\chi}_2^0}$   &     \text{183.36 GeV} & \text{0.08 GeV  }\\ %    # ~chi_20
$m_{\tilde{\chi}_1^{\pm}}$ &     \text{181.85 GeV} & \text{0.55 GeV  }\\ %    # ~chi_1+
$m_{\tilde{\chi}_2^{\pm}}$ &     \text{380.4 GeV} & \text{3.0 GeV   }\\ %    # ~chi_2+

%\hline
%\end{tabular}
%}
%\end{center}
%\end{table}
%\begin{table}
%\begin{center}
%{
%\begin{tabular}{|l|| c |c|}
%\hline 
%Measurement & Value & Uncertainty \\ 
%\hline\hline
$\sigma_+$ ( $\text{e}^+\text{e}^- \rightarrow \tilde{\chi}_1^0 \tilde{\chi}_2^0$)  &  \text{20.9 fb}&  \text{1.8  fb}  \\ %alias 1
$\sigma_+$ ( $\text{e}^+\text{e}^- \rightarrow \tilde{\chi}_2^0 \tilde{\chi}_2^0$)  &  \text{17.3 fb}&  \text{1.8  fb}  \\ %alias 2
$\sigma_+$ ( $\text{e}^+\text{e}^- \rightarrow \tilde{\text{e}}_L \tilde{\text{e}}_L$)  &  \text{156.3 fb}&  \text{3.0  fb}  \\ %alias 3
$\sigma_+$ ( $\text{e}^+\text{e}^- \rightarrow \tilde{\mu}_L \tilde{\mu}_L$)              &  \text{27.0   fb}&  \text{2.9  fb}  \\ %alias 4
$\sigma_+$ ( $\text{e}^+\text{e}^- \rightarrow \tilde{\tau}_1 \tilde{\tau}_1$)             &  \text{28.8 fb}&  \text{2.9  fb}  \\ %alias 5
$\sigma_+$ ( $\text{e}^+\text{e}^- \rightarrow \tilde{\chi}_1^{\pm} \tilde{\chi}_1^{\mp}$)      &  \text{43.5 fb}&  \text{0.9  fb}  \\ %alias 6
$\sigma_+$ ( $\text{e}^+\text{e}^- \rightarrow$ Z $\text{h}^0$)     &  \text{11.14 fb}&  \text{0.21 fb}  \\ %  alias 7
$\sigma_-$ ( $\text{e}^+\text{e}^- \rightarrow \tilde{\chi}_1^{\pm} \tilde{\chi}_1^{\mp}$)    &  \text{97.6 fb}&  \text{3.3  fb}  \\ %alias 8
$\sigma_-$ ( $\text{e}^+\text{e}^- \rightarrow \tilde{\chi}_1^0 \tilde{\chi}_2^0$)&  \text{40.2 fb}&  \text{1.8  fb}  \\ %alias 9
$\sigma_-$ ( $\text{e}^+\text{e}^- \rightarrow \tilde{\chi}_2^0 \tilde{\chi}_2^0$)&  \text{38.8 fb}&  \text{1.8  fb}  \\ %alias 10
$\sigma_-$ ( $\text{e}^+\text{e}^- \rightarrow \tilde{\text{e}}_L \tilde{\text{e}}_L$)&  \text{74.1 fb}&  \text{3.0  fb}  \\ %alias 11
$\sigma_-$ ( $\text{e}^+\text{e}^- \rightarrow \tilde{\text{e}}_L \tilde{\text{e}}_R$)&  \text{169.0 fb}&  \text{3.0  fb}  \\ %alias 12
$\sigma_-$ ( $\text{e}^+\text{e}^- \rightarrow \tilde{\text{e}}_R \tilde{\text{e}}_R$)&  \text{14.4 fb}&  \text{1.0  fb}  \\ %alias 13
$\sigma_-$ ( $\text{e}^+\text{e}^- \rightarrow \tilde{\mu}_L \tilde{\mu}_L$)            &  \text{16.6 fb}&  \text{1.5  fb}  \\ %alias 14
$\sigma_-$ ( $\text{e}^+\text{e}^- \rightarrow \tilde{\tau}_1 \tilde{\tau}_1$)           &  \text{18.8 fb}&  \text{1.5  fb}  \\ %alias 15

BR ( $\text{h}^0 \rightarrow \text{b}\bar{\text{b}}$ )                       &  \text{0.83 } & \text{0.01} \\ %     alias 1
BR ( $\text{h}^0 \rightarrow \text{c}\bar{\text{c}}$)                        &  \text{0.04} & \text{0.01} \\ %     alias 2
BR ( $\text{h}^0 \rightarrow \tau^+ \tau^-$ )                            &  \text{0.13 } & \text{0.01}  \\ %    alias 3
\hline
\end{tabular}
}
\end{center}
\caption{The simulated LHC and LC measurements for the considered SUSY
  scenario. Standard Model parameters and squark masses are not listed.
 The cross sections correspond to a centre-of-mass energy of
  $\sqrt{s} = 500$ GeV. The electron and positron polarisations are indicated
  by subscript: ``$+$'' for $P_{\text{e}^-} = 0.8$, $P_{\text{e}^+} = 0.6$ and
  ``$-$'' for $P_{\text{e}^-} = -0.8$, $P_{\text{e}^+} = -0.6$.}
\label{tbl:obs}
\end{table}
Now, three fits have been performed: One, with only the observables from
Table~\ref{tbl:obs} without the branching ratio as an included measurement.
The second one includes $BR({\tilde{\chi}_1^\pm}\rightarrow {\tilde{\chi}_1^0}
W^\pm) = 33.4\%$, which is the numerical value obtained with {\it SPHENO},
together with a relative measurement error of 7.5\% as the result for
${\sqrt{s_{ee}}=500 \gev}$.  The third fit is similar but with an error of
3.5\% obtained in the as the result for the ${\sqrt{s_{ee}}=600 \gev}$ case.
Table~\ref{tbl:fitres} shows the fitted parameters and the uncertainties
obtained from the three fits. Because we were just interested in the
final errors, we simply used the actual input values of the parameters as start
values for the fit. 
In terms of precision, many parameters are not
influenced significantly. However the uncertainties on the parameters determining the chargino and
neutralino mixing matrices, especially $\tan\beta$, and on $X_\tau$ improve,
when the branching ratio is added as a measured observable. For $\tan\beta$
the relative error improves by a factor of 2 for ${\Delta BR}/{BR} = 3.5\%$.
The errors for the stau masses $m_{\tilde{\tau}_R}$, $m_{\tilde{\tau}_L}$ also
get better by roughly a factor of 2. The errors of some other parameters
(e.g. $M_{\tilde{e}_R}$) might improve a little because of an overall
correlation among all fitted parameters. The improper decrease of
precision on $\mu$ and $M_2$ is due to a slightly unstable fit.
It should, however, be noted that up to
now no observables sensitive to the decay modes of the superpartners have been
studied in $e^+ e^-$.
\begin{table}[hbt]
\begin{center}
{\footnotesize
\renewcommand{\arraystretch}{1.1}
\begin{tabular}{|l|c||l|c||l|c|}
\hline
Parameter & Value (\gev)& Parameter & Value (\gev)& Parameter & Value (\gev)\\
\hline\hline
$X_t$ & -535.09 & $X_b$ &-3972.09 & $M_3$ & 579.42 \\ 
$m_{\tilde{d}_R}$ & 525.15 & $m_{\tilde{s}_R}$ & 525.15 & $m_{\tilde{b}_R}$ & 522.65\\
$m_{\tilde{u}_R}$ & 527.24 & $m_{\tilde{c}_R}$ & 527.24 & $m_{\tilde{t}_R}$ & 423.98\\
$m_{\tilde{u}_L}$ & 544.21 & $m_{\tilde{c}_L}$ & 544.21 & $m_{\tilde{t}_L}$ & 497.43\\
$m_t$ & 174.3 & $m_b$ & 4.2 & $m_c$ & 1.2\\
\hline
\end{tabular}
 }
\end{center}
\caption{The fixed parameters and their input values.}
\label{tbl:fixed}
\end{table}
\begin{table}[hbt]
\begin{center}
\renewcommand{\arraystretch}{1.15}
\begin{tabular}{|l||c|c|c|c|}
\hline
& & \multicolumn{3}{c|}{uncertainty}\\
Parameter & Value (\gev)& without $BR$ & $\frac{\Delta BR}{BR} = 7.5\%$ & $\frac{\Delta BR}{BR} = 3.5\%$\\
\hline\hline
$\tan\beta$ & 9.00 & 22\% & 16\% & 10\%\\
$X_\tau$ & -3457.5 & 19\% & 7\% & 6\% \\
$\mu$ & 355.96 & 1.2\% & 1.4\% & 1.0\% \\
$M_1$ & 99.54 & 0.3\% & 0.3\% & 0.2\%\\
$M_2$ & 192.57 & 0.4\% & 0.6\% & 0.3\%\\
$m_{A_0}$ & 406.59 & 0.2\% & 0.2\% & 0.2\%\\
$M_{\tilde{\tau}_R}$ & 157.31 & 1.3\% & 0.5\% & 0.5\%\\
$M_{\tilde{\tau}_L}$ & 212.28 & 1.0\% & 0.6\% & 0.6\%\\
$M_{\tilde{\mu}_R}$ & 159.41 & 0.15\% & 0.15\% & 0.15\%\\
$M_{\tilde{\mu}_L}$ & 213.04 & 0.3\% & 0.3\% & 0.3\%\\
$M_{\tilde{e}_R}$ & 159.41 & 0.05\% & 0.05\% & 0.04\%\\
$M_{\tilde{e}_L}$ & 213.04 & 0.10\% & 0.09\% & 0.09\%\\

\hline
\end{tabular}
\end{center}
\caption{The fitted parameters and the uncertainties obtained in the
  fits. The second column lists the input values.}
\label{tbl:fitres}
\end{table}

%%%%%%%%%%%%%%%%%%%%%%%%%%%%%%%%%%%%%%%%%%%%%%%%%%%%%%%%%%%%%%%%%%%%%%%
\section{Conclusions}
%%%%%%%%%%%%%%%%%%%%%%%%%%%%%%%%%%%%%%%%%%%%%%%%%%%%%%%%%%%%%%%%%%%%%%%

A future photon collider provides the opportunity to measure the branching
ratio of the chargino decay ${\tilde{\chi}_1^\pm}\rightarrow
{\tilde{\chi}_1^0} W^\pm$ directly. Considering a mSUGRA scenario similar to
SPS1a, this Monte Carlo study showed that a statistical error for the
branching ratio of ${\Delta BR}/{BR} = 3.5\%$ (7.5\%) for an electron
centre-of-mass energy of $\sqrt{s} = 600$ GeV ($\sqrt{s} = 500$ GeV) can be
obtained. Such a measurement would improve the precision of a global MSSM
parameter fit.

%%%%%%%%%%%%%%%%%%%%%%%%%%%%%%%%%%%%%%%%%%%%%%%%%%%%%%%%%%%%%%%%%%%%%%%
\section*{Acknowledgements}
%%%%%%%%%%%%%%%%%%%%%%%%%%%%%%%%%%%%%%%%%%%%%%%%%%%%%%%%%%%%%%%%%%%%%%%

We would like to thank the creators of {\it SHERPA} especially Frank Krauss,
Andreas Sch\"alicke, Steffen Schumann and Tanju Gleisberg for their dedicated
help and support. We thank Philip Bechtle and Peter Wienemann for 
the important assistance in the usage of {\it Fittino}. 
We also thank Hanna Nowak and Sabine Riemann for many helpful discussions.

\boldmath

\end{document}